\documentclass{emulateapj}

\usepackage{amssymb}
\usepackage{amsmath}
\usepackage{graphicx}
\usepackage{natbib}
\usepackage{txfonts}
\usepackage[colorlinks=true,citecolor=blue,linkcolor=blue]{hyperref}
\usepackage{microtype}

\usepackage{color}

\shortauthors{Pach\'on et al.}

\bibpunct{(}{)}{;}{a}{}{,} 

\def\Xint#1{\mathchoice
{\XXint\displaystyle\textstyle{#1}}%
{\XXint\textstyle\scriptstyle{#1}}%
{\XXint\scriptstyle\scriptscriptstyle{#1}}%
{\XXint\scriptscriptstyle\scriptscriptstyle{#1}}%
\!\int}
\def\XXint#1#2#3{{\setbox0=\hbox{$#1{#2#3}{\int}$}
\vcenter{\hbox{$#2#3$}}\kern-.5\wd0}}

\def\dashint{\Xint-}
\def\mnras{Mon. Not. Roy. Astr. Soc.}
\def\apj{Astrophys. J.}
\def\apjl{Astrophys. J. Lett.}
\def\apjs{Astrophys. J. Suppl. Ser.}

\def\aap{Astron. Astrophys.}

\begin{document}


\title{On the relativistic precession and oscillation frequencies of test particles around rapidly rotating compact stars
}

\author{Leonardo~A.~Pach\'on}
\affil{Instituto de F\'isica, Universidad de Antioquia, AA 1226 Medell\'in, Colombia}

\author{Jorge A.~Rueda}
\affil{Dipartimento di Fisica and ICRA, Sapienza Universit\`a di Roma, P.le Aldo Moro 5, I-00185 
Rome, Italy\\ICRANet, P.zza della Repubblica 10, I--65122 Pescara, Italy}

\and

\author{C\'esar.~A.~Valenzuela-Toledo}
\affil{Departamento de F\'isica, Universidad del Valle, A.A. 25360, Santiago de Cali, Colombia}

\altaffiltext{1}{leonardo.pachon@fisica.udea.edu.co}
\altaffiltext{2}{jorge.rueda@icra.it}
\altaffiltext{3}{cesar.valenzuela@correounivalle.edu.co}

\begin{abstract}
Whether analytic exact vacuum(electrovacuum) solutions of the Einstein(Einstein-Maxwell) field equations can accurately describe or not the exterior spacetime of compact stars remains still an interesting open question in Relativistic Astrophysics. As an attempt to establish their level of accuracy, the radii of the Innermost Stable Circular Orbits (ISCOs) of test particles given by analytic exterior spacetime geometries have been compared with the ones given by numerical solutions for neutron stars (NSs) obeying a realistic equation of state (EoS). It has been so shown that the six-parametric solution of Pach\'on, Rueda, and Sanabria (2006) (hereafter PRS) is more accurate to describe the NS ISCO radii than other analytic models. We propose here an additional test of accuracy for analytic exterior geometries based on the comparison of orbital frequencies of neutral test particles. We compute the Keplerian, frame-dragging, as well as the precession and oscillation frequencies of the radial and vertical motions of neutral test particles for the Kerr and PRS geometries; then we compare them with the numerical values obtained by Morsink and Stella (1999) for realistic NSs. We identify the role of high-order multipole moments such as the mass quadrupole and current octupole in the determination of the orbital frequencies especially in the rapid rotation regime. The results of this work are relevant to cast a separatrix between black hole (BH) and NS signatures as well as probe the nuclear matter EoS and NS parameters from the Quasi-Periodic Oscillations (QPOs) observed in Low Mass X-Ray Binaries.
\end{abstract}


\keywords{Exact solutions Einstein-Maxwell equations -- Relativistic precession frequencies}



\section{Introduction}
One of the greatest challenges of the general theory of re\-la\-tivity has been the construction of solutions to the Einstein-Maxwell field equations representing the gravitational field of compact stars such as neutron stars (NSs). Stationary axially symmetric spacetimes satisfy basic properties one expects for rotating objects, namely time symmetry and reflection symmetry with respect to the rotation axis \citep[see e.g.][]{2006CQGra..23.3251P}. The simplest stationary axially symmetric exact exterior vacuum solution describing a rotating configuration is the well-known Kerr metric \citep{1963PhRvL..11..237K}. The Kerr metric is fully described by two free parameters: the mass $M$ and the angular momentum $J$ of the object. However, it is known from numerical models that the quadrupole moment of rotating NSs deviates considerably from the one given by the Kerr solution $Q_{\rm Kerr}=-J^2/(M c^2)$ \citep[see e.g.][for details]{1999ApJ...512..282L}. 

In the mean time, a considerable number of analytic exterior solutions with a more complex multipolar structure than the one of the Kerr solution have been developed \citep[see e.g.][]{1995JMP....36.3063M,2000PhRvD..62d4048M,exactsolbook}. Whether analytic exterior solutions are accurate or not to describe the gravitational field of compact stars is an interesting and very active topic of research \citep[see e.g.][and references therein]{2002MNRAS.336..831S,2004MNRAS.350.1416B,2006PhRvD..73j4038P}.

The accuracy of analytic solutions to describe the exterior geometry of a realistic rotating compact star has been tested by comparing physical properties, e.g. the radius of the Innermost Stable Circular Orbit (ISCO) on the equatorial plane and the gravitational redshift \citep[see][for details]{1998AstL...24..774S,2004MNRAS.350.1416B,2006PhRvD..73j4038P}. In order to do such a comparison, the free parameters (i.e. the lowest multipole moments) of the analytic exterior spacetime, are fixed to the corresponding lowest multipole moments given by numerical interior solutions of the Einstein equations, for NS realistic models 
\citep[see e.g.][]{2004MNRAS.350.1416B}.

Following such a procedure, the solution of \citet{2000PhRvD..62d4048M} has been compared by \citet{2002MNRAS.336..831S} and by \citet{2004MNRAS.350.1416B} with the numerical solutions for NSs calculated by \citet{1994ApJ...424..823C} and with those derived by \citet{2004MNRAS.350.1416B}, respectively. However, being a generalization of the solution of \citet{1972PhRvL..29.1344T}, it cannot describe slowly rotating compact stars 
\citep[see e.g.][]{2004MNRAS.350.1416B}, but the dynamics of astrophysical objects with anisotropic stresses \citep[see][for details]{2007PhRvD..75b3008D}.

Following a similar procedure, based on tests of the ISCOs radii on the equatorial plane of the rotating neutron stars obtained by \cite{2004MNRAS.350.1416B}, it has been shown that the six-parametric solution of \cite{2006PhRvD..73j4038P} (hereafter PRS solution, see Sec.~\ref{sec:2} for details) is more accurate than the model of \cite{2000PhRvD..62d4048M}. In addition, being a generalization of the Kerr solution, this solution can be used for arbitrary rotation rates. 

Besides the ISCOs radii, there are additional physical properties that can be computed with analytic and numerical models and thus useful to compare and contrast the accuracy of analytic exact models. The aim of this article is to analyze the properties of orbital frequencies of neutral test particles in the PRS and in the Kerr geometries with especial focus on the Keplerian $\nu_{\rm K}$, frame-dragging (Lense-Thirring) $\nu_{\rm LT}$, as well as the precession(oscillation) frequencies of the radial and vertical motions, $\nu^{\rm P}_{\rho}$($\nu^{\rm OS}_{\rho}$) and $\nu^{\rm P}_{z}$($\nu^{\rm OS}_{z}$), respectively.

The relevance of these frequencies relies on the fact that they are often invoked to explain the Quasi-Periodic Oscillations (QPOs) observed in some relativistic astrophysical systems such as Low Mass X-ray Binaries (LMXBs), binary systems harboring either a NS or a black hole (BH) accreting matter from a companion star. For instance, within the Relativistic Precession Model (RPM) introduced by \cite{1998ApJ...492L..59S,1999ApJ...513..827M,1999ApJ...524L..63S,1999PhRvL..82...17S}, the kHz QPOs are interpreted as a direct manifestation of the modes of relativistic epicyclic motion of blobs arising at various radii $r$ in the inner parts of the accretion disk around the compact object (see Sec.~\ref{sec:RPM}, for details).

In addition to the RPM, the Keplerian, precession and oscillation frequencies are used in other 
QPO theoretical models \citep[see e.g.][for a recent comparison of the existing models]{2011ApJ...726...74L}. 
Due to the influence of general relativistic effects in the determination of such frequencies, an 
observational confirmation of any of the models might lead to an outstanding test of general 
relativity in the strong field regime. In this line, it is of interest to compare and contrast the orbital 
frequencies given by the Kerr solution and by the PRS solution (see Sec.~\ref{sec:3}), which help 
to establish the differences between possible BH and NS signatures. We emphasize in this article 
the major role of the quadrupole moment as well as of the octupole moment of the object, whose 
possible measurement can be used as a tool to test the no-hair theorem of BHs 
\citep[see e.g.][]{2011ApJ...726...11J} and to discriminate between the different theoretical models 
proposed to explain the physics at interior and exterior of the Neutron Stars. 
Additionally, in the case of NSs, the interpretation of QPOs as the 
manifestation of orbital motion frequencies might lead to crucial information of the NS parameters 
such as mass, angular momentum \citep[see e.g.][]{1998ApJ...492L..59S,2010ApJ...714..748T}, 
and quadrupole moment \citep[see e.g.][]{1999ApJ...513..827M}. These parameters reveal, at the 
same time, invaluable information about the EoS of nuclear matter. 

The article is organized as follows. In Sec.~\ref{sec:2} we recall the properties of the PRS solution. 
The computation of the orbital frequencies as well as the comparison of their features in the Kerr 
and in the PRS spacetimes, is shown in Sec.~\ref{sec:3}. 
In Sec.~\ref{sec:4} we study the accuracy of the analytic formulas of the periastron and nodal 
frequencies derived by \cite{1995PhRvD..52.5707R} for stationary axially symmetric spacetimes. 
\textcolor{black}{In Sections 5 and 6 we discuss the accuracy of the PRS solution in describing
the frequencies of realistic NS models and its relevance in the Relativistic Precession Model, 
respectively.} The conclusions of this work and a discussion on possible additional effects to be accounted for in the determination of the orbital frequencies, e.g. the effect of magnetic dipole moment, are outlined in Sec.~\ref{sec:6}.

\section[]{The PRS analytic exact solution}\label{sec:2}
We first recall the PRS analytic model \citep{2006PhRvD..73j4038P}, for the exterior gravitational field of a compact object\footnote{Mathematica 8.0 scripts with the solution, some limiting cases as well as the the calculations presented in this paper are available at 
\href{http://www.chem.utoronto.ca/~lpachon/scripts/nstars}
{http://www.chem.utoronto.ca/$\sim$lpachon/scripts/nstars}}. 
In the stationary axisymmetric case, the simplest form of the metric can be written as \citep{1953AnP...447..309P}
\begin{equation}
\label{Papapetrou} 
ds^2=-f(dt-\omega d\phi)^2+f^{-1}\left[e^{2\gamma} (d\rho^2+dz^2)+\rho^2d\phi^2 \right]\, ,
\end{equation}
where $f$, $\omega$ and $\gamma$ are functions of the quasi--cylindrical Weyl coordinates $(\rho,z)$. Thus, the components of the metric tensor $g_{\mu\nu}$ are
\begin{align}
g_{\phi \phi} &= \frac{\rho^2}{f(\rho,z)} - f(\rho,z) \omega(\rho,z)^2,
\\
g_{tt} &= -f(\rho,z),
\\
g_{t\phi} &= f(\rho,z) \omega(\rho,z),
\\
g_{zz} &= g_{\rho \rho} = \frac{{\rm e}^{2\gamma(\rho,z)}}{f(\rho,z)}
= \frac{1}{g^{zz}} =  \frac{1}{g^{\rho\rho}}.
\end{align}

Using the above line element, the Einstein-Maxwell equations can be reformulated, via 
Ernst's procedure in terms of two complex potentials ${\cal E}(\rho,z)$ and $\Phi(\rho,z)$ 
\citep{1968PhRv..167.1175E,1968PhRv..172.1850E}. By means of Sibgatullin's integral 
method \citep{1991owsg.book.....S,1993CQGra..10.1383M} this system of equations can 
be solved va
\begin{align}
\label{Ernst1}
{\cal E}(z,\rho)&=\int\limits_{-1}^1\frac{d\sigma}{\pi}
\frac{e(\xi)\mu(\sigma)}{\sqrt{1-\sigma^2}}, \\
\Phi(z,\rho)&=\int\limits_{-1}^1 \frac{d\sigma}{\pi}
\frac{f(\xi)\mu(\sigma)}{\sqrt{1-\sigma^2}},
\end{align}
where $e(z):={\cal E}(z,\rho=0)$ and \mbox{$f(z):=\Phi(z,\rho=0)$}. The unknown function 
$\mu(\sigma)$ must satisfy the singular integral equation
\begin{equation}
\dashint_{-1}^{1}\frac{\mu(\sigma)[e(\xi)+\tilde
e(\eta)+2f(\xi)\tilde
f(\eta)]d\sigma}{(\sigma-\tau)\sqrt{1-\sigma^2}}=0
\end{equation}
and the normalizing condition
\begin{equation}
\int_{-1}^1\frac{\mu(\sigma)d\sigma}{\sqrt{1-\sigma^2}}=\pi,
\end{equation}
where $\xi=z+i\rho\sigma$, $\eta=z+i\rho\tau$, $\rho$ and $z$ being the Weyl-Papapetrou 
quasi--cylindrical coordinates, $\sigma, \tau\in[-1,1]$, $\tilde e(\eta):=\overline{e(\bar\eta)}$, 
$\tilde f(\eta):=\overline{f(\bar\eta)}$ and the overbar stands for complex conjugation. In 
\citep{2006PhRvD..73j4038P}, the Ernst potentials were chosen as 
\begin{align}
\label{Potenciales eje}
\begin{split}
e(z) = \frac{z^3-z^2(m+\textrm{i}a)-kz+\textrm{i}s}{z^3+z^2(m-\textrm{i}a)-kz+\textrm{i}s}\, ,
\\
f(z) = \frac{q z^2 + \textrm{i}\mu z}{z^3+z^2(m-\textrm{i}a)-kz+\textrm{i}s}\, .
\end{split}
\end{align}

We calculate the multipole moments following the procedure of \cite{1990CQGra...7.1819H}. 
We denote the mass multipoles by $M_i$ while, the current (rotation) multipoles, by $S_i$. 
The electric multipoles are denoted by $Q_i$ and the magnetic ones by ${\cal B}_i$. 
Thus, for the PRS solution we have
\begin{align}
\begin{split}
\label{multipolosP}
M_0 &= m\, ,\quad M_2 = m k - m a^2\, ,\quad \ldots \\
S_1 &= m a\, ,\quad S_3 = - m a^3 + 2 m a k  - m s\, ,\quad \ldots \\
\end{split}
\\
\begin{split}
\label{multipolosQ}
Q_0 &= q\, ,\quad Q_2 = - a^2 q - a \mu + k q\, ,\quad \ldots \\
{\cal B}_1 &= \mu + a q\, ,\quad {\cal B}_3 = - a^2 \mu + \mu k - a^3 q + 2 a k q - q s\, ,\quad \ldots
\end{split}
\end{align}
This allows us to identify $m$ as the total mass, $a$ as the total angular moment per unit 
mass ($a=J/m$, being $J$ the total angular moment); while $k$, $s$, $q$ and $\mu$ are 
associated to the mass-quadrupole moment $M_2$, current octupole $S_3$, electric charge 
and magnetic dipole, respectively.

The potentials (\ref{Potenciales eje}) can be written in an alternative way, we mean
\begin{equation}
e(z)=1+\displaystyle \sum_{i=3}^{3} \frac{e_{i}}{z-\beta_{i}}\,
,\qquad f(z)=\displaystyle \sum_{i=3}^{3}
\frac{f_{i}}{z-\beta_{i}}\, ,
\end{equation}
with
\begin{align}
e_{j}&= (-1)^{j}\frac{2 m
\beta^{2}_{j}}{(\beta_{j}-\beta_{k})(\beta_{j}-\beta_{i})}\, ,
\\
\label{equ:f_i}
f_{j}&= (-1)^{j+1}\frac{i \mu + d\beta_{j}
}{(\beta_{j}-\beta_{k})(\beta_{j}-\beta_{i})}\, , \quad i,k \neq j\, .
\end{align}
Then, using Eqs.~(\ref{Ernst1}) and (\ref{Potenciales eje}), we obtain the Ernst potentials
\begin{equation}\label{potenciales_ernst}
{\cal E}=\frac{A + B }{A - B}\, , \qquad \Phi=\frac{C}{A - B}\, ,
\end{equation}
and the metric functions in the whole spacetime
\begin{align}\label{eq:metricfuncs}
f&=\frac{A \bar{A}-B \bar{B} + C \bar{C}}{( A -
B)(\bar{A}-\bar{B})}\, ,\quad e^{2\gamma}=\frac{A \bar{A} -B
\bar{B} + C \bar{C}}{\displaystyle{K \bar{K}\prod_{n=1}^{6}r_n}}\, ,\\
\omega &= \frac{{\rm Im}[(A + B)\bar{H}-(\bar{A} + \bar{B})G - C
\bar{I}]}{A \bar{A} - B \bar{B} + C \bar{C}}\, ,
\end{align}
where the functions $A$, $B$, $C$, $H$, $G$, $K$, and $I$ can be found in the Appendix 
\ref{app:metricfuncs}.

The PRS electrovacuum exact solution belongs to the extended $N$-soliton solution of the 
Einstein-Maxwell equations derived by \cite{1995PhRvD..51.4192R}, in the particular case 
$N=3$. 
In addition, the functional form of the metric functions resembles the one derived previously 
by \cite{1999CQGra..16.3725B}. 
Besides the limiting cases discussed in \cite{2006PhRvD..73j4038P} it is worth mentioning that, 
in the vacuum case $q=0$ and $\mu=0$, for $s=0$ this solution reduces to the solution of 
\cite{1995JMP....36.3063M} under the same physical conditions, namely $q=0$, $c=0$ and 
$b=0$ in \cite{1995JMP....36.3063M}.

\section{Orbital Motion Frequencies on the Equatorial Plane}\label{sec:3}

Although for the case of compact stars contributions from the magnetic field could be relevant 
\citep[see e.g.][]{2010CQGra..27d5001B,2010PhRvD..82l4014S,2012CQGra..29f5012B}, we 
focus in this work on the frequencies of neutral particles orbiting a neutral compact object. 
We calculate here the Keplerian 
$\nu_{\rm K}=\Omega_{\rm K}/(2 \pi)$, frame-dragging (Lense-Thirring) $\nu_{\rm LT}=\Omega_{\rm LT}/(2 \pi)$, 
radial oscillation and precession, $\nu^{\rm OS}_{\rho}=\Omega^{\rm OS}_{\rho}/(2 \pi)$ and 
$\nu^{\rm P}_{\rho}=\Omega^{\rm P}_{\rho}/(2 \pi)$, and vertical oscillation and precession 
frequencies, $\nu^{\rm OS}_{z}=\Omega^{\rm OS}_{z}/(2 \pi)$ and 
$\nu^{\rm P}_{\rho}=\Omega^{\rm P}_{\rho}/(2 \pi)$, respectively.

\begin{figure}
\centering
\includegraphics[width=\columnwidth,clip]{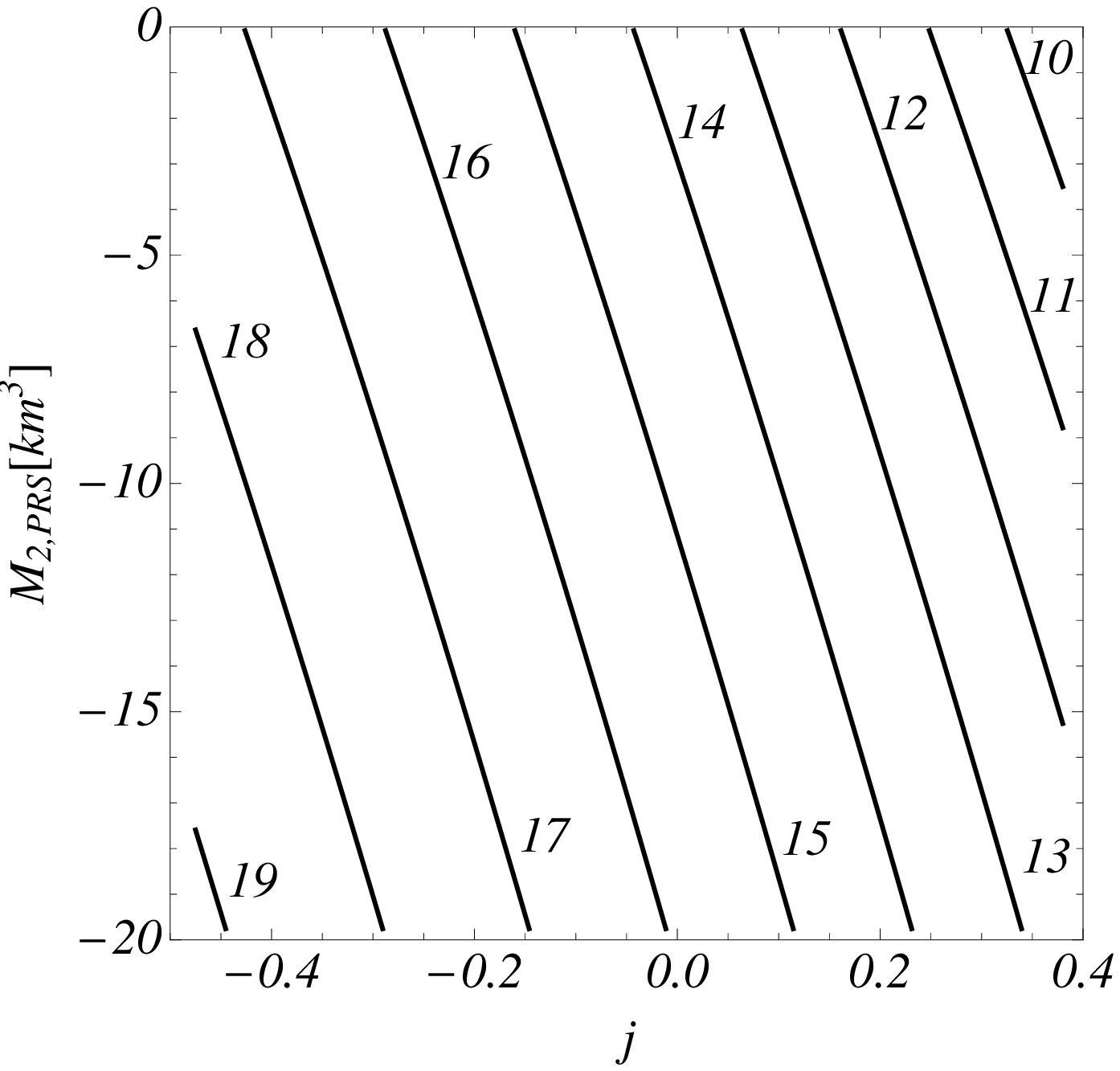}\\
\includegraphics[width=\columnwidth,clip]{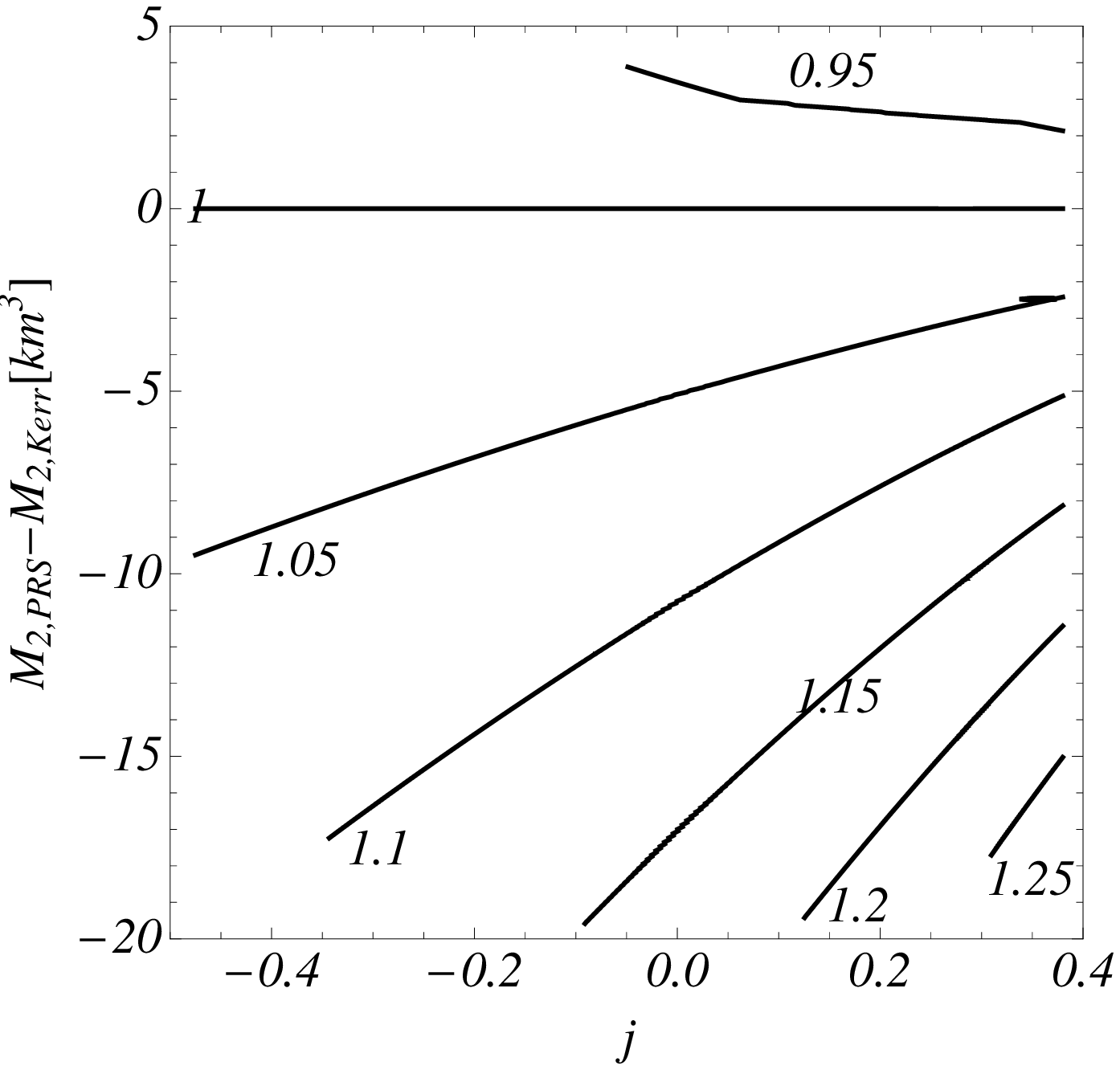}
\caption{Upper panel: Contours of constant ISCO radius as a function of the dimensionless angular momentum parameter $j=J/M^2_0$ and the quadrupole moment $M_2$ for the PRS solution, for a compact object with mass $M_0=m=1.88 M_{\odot}=2.78$ km. Contours are labeled by the corresponding the value of the ISCO radius in km. 
\textcolor{black}{Negative values of $j$ depict the counter-rotating case and negative values of the quadrupole moment $M_2$ correspond to oblate configurations. The values of $M_2$ are in the range $0\leq M_2\leq 20$ km$^3$ that corresponds in CGS units to $0\leq M_2\leq -2.7\times10^{44}$ g cm$^2$, which covers the typical range of fast rotating NSs.} Lower panel: Contours of constant ratio $r_{\rm ISCO,PRS}/r_{\rm ISCO,Kerr}$ as a function of $j$ 
and the difference $M_{2,\rm PRS}-M_{2,\rm Kerr}$. 
\textcolor{black}{The quadrupole moment difference is comprised in the range $-2.7\times10^{44} \leq M_2 \leq 6.8\times10^{43}$ g cm$^2$.}
}
\label{fig:plotrome}
\end{figure}

The geodesic motion of test particles along the radial coordinate, on the equatorial plane $z=0$, 
is governed by the effective potential  \citep[see e.g.][]{1995PhRvD..52.5707R}
\begin{equation}
\label{equ:EffPot}
V(\rho) = 1 - \frac{E^2 g_{\phi\phi} + 2 E L g_{t\phi} + L^2 g_{tt}}{g_{t\phi}^2 - g_{tt} g_{\phi\phi}},
\end{equation}
where, for circular orbits, the energy $E$ and angular momentum $L$ are determined by the 
conditions $V=0$ and $\mathrm{d}V/\mathrm{d}\rho = 0$ (see Eqs.~\ref{equ:E}--\ref{equ:L}). 
The frequencies at the ISCO's location (determined by the additional condition  
$\mathrm{d}^2V/\mathrm{d}\rho^2 = 0$) are of particular interest. 
Thus, before starting the discussion of the frequencies, it is important to explore the ISCO parametric dependence. 
We report here, as standard in the literature, the \emph{physical} ISCO radius given by $\sqrt{g_{\phi\phi}}$ evaluated at the root of Eq.~(\ref{equ:EffPot}) that gives the coordinate ISCO radius. 
In the upper panel of Fig.~\ref{fig:plotrome} we plotted contours of constant ISCO radii as a function of the dimensionless angular momentum parameter $j=J/M^2_0$ and the star quadrupole moment $M_2$, for the PRS solution. 
\textcolor{black}{The use of the dimensionless parameter $j$ in the horizontal axis allows to, qualitatively, relate deviations of the contour lines from vertical lines to the influence of the quadrupole moment.} 
We can see that the ISCO radius decreases for increasing $j$ and decreasing $M_2$. 
\textcolor{black}{A quantitative measurement of this influence could be derived from the effective slope of the contour lines. 
We are interested in the comparison with the Kerr geometry, so in the lower panel,} we plotted contours of constant ratio 
$r_{\rm ISCO,PRS}/r_{\rm ISCO, Kerr}$ as a function of $j$ and the difference between the quadrupole moment of the PRS solution $M_{2,\rm PRS}$ and the Kerr quadrupole $M_{2,\rm Kerr} = - m a^2$, i.e.~$M_{2,\rm PRS}-M_{2,\rm Kerr} = M_{2,\rm PRS}+m a^2 = m k$, see Eq.~(\ref{multipolosP}). 
Deviations from the Kerr geometry are evident. 
Negative values of the angular momentum correspond to the radii of the counter-rotating orbits obtained here 
through the change $g_{t\phi} \rightarrow -g_{t\phi}$ (see discussion below).

We stress that the accuracy of the PRS solution for describing the ISCO radius of \emph{realistic} 
NSs was already shown to be higher with respect to other analytic models \citep[see][for details]{2006PhRvD..73j4038P}. 
In Table~\ref{tab:ISCO} we compare the ISCO radius for two rapidly rotating NS, models 20 and 26, of Table VI of \cite{PA12} for the EoS L. The lowest multipole moments of the analytic models are fixed to the numerical values obtained by \cite{PA12}. 
In the case of the Kerr solution, only $M_0$ and $J$ can be fixed, while $M_2$, and $S_3$ have values that depend on $M_0$ and $J$ and therefore cannot be fixed. 
For the PRS solution with $s=0$, $M_0$, $J$ and $M_2$ can be fixed while $S_3$ remains induced by the lower moments. We present also the ISCO radius obtained by fixing $M_0$, $J$, $M_2$, as well as $S_3$ in the PRS analytic exact model. 
\begin{table}
\centering
\begin{tabular}{cccccc}
& $R_{\rm N}$[km] & $R_{\rm SS}$[km] & $R_{\rm Kerr}$[km]
& $R_{{\rm PRS},s=0}$[km] & $R_{\rm PRS}$[km]\\
\tableline
M20& 19.81&13.39&16.14&19.28& 18.99\\
M26& 19.87&17.16&15.94&19.65 & 19.54\\
\tableline
\end{tabular}
\caption{Comparison of the ISCO radius for the selected NS models 20 and 26 of Table VI of \cite{PA12} for the EoS L. Model 20: $M_0=4.167$ km ($2.82 M_\odot$), $j=J/M^2_0 = 0.70$, $M_2 = -79.8$ km$^3$ ($-1.08\times 10^{45}$ g cm$^2$) and $S_3=-401.0$ km$^4$. Model 26: $M_0=4.36$ km ($2.95 M_\odot$), $j=J/M^2_0 = 0.56$, $M_2 = -45.2$ km$^3$ ($-6.10\times 10^{44}$ g cm$^2$) and $S_3=-170.0$ km$^4$. The subscript N stands for the numerical calculation of \cite{PA12} and SS stands for the \cite{SS98} approximated ISCO radius expression.}
\label{tab:ISCO}
\end{table}

In Figs.~\ref{fig:plotrome}--\ref{fig:plotomegaLT}, we have fixed as an example $M_0=m=1.88 M_{\odot}=2.78$ km, and $s=0$. We recall that the quadrupole moment in the geometric units used here (km$^3$) is related to the one in CGS units by $M^{\rm CGS}_2=(10^{15} c^2/G) M^{\rm geo}_2=1.35 \times 10^{43} (M^{\rm geo}_2/{\rm km}^3)$ g cm$^2$, and the mass of the Sun is $M^{\rm geo}_\odot=1.477$ km. The dimensionless angular momentum $j$ is obtained from the CGS values of $J$ and $M_0$ as $j=c J/(G M^2_0)$.

It is appropriate to compare the range of values of $j=J/M^2_0$ and $M_2$ used in 
Figs.~\ref{fig:plotrome}--\ref{fig:plotomegaLT} with typical values of a NS. 
For the used mass $M_0=1.88 M_{\odot}$, \cite{1999ApJ...513..827M} obtained a 
quadrupole moment $M_2=-5.3\times 10^{43}$ g cm$^2=3.93$ km$^3$, with the 
latter value in geometric units, for a NS of angular rotation frequency $\nu_s=290$~Hz 
(rotation period of $3.45$ milliseconds), corresponding to a dimensionless angular 
momentum $j=J/M^2_0=0.19$, for the EoS L. For a fixed mass the quadrupole moment 
is an increasing function of $j$ because an increasing of the angular momentum at fixed 
mass results in an increasing of the oblateness (eccentricity) of the star, and so the 
quadrupole moment. 
Based on this fact, it is clear that not all the values of the $(M_2,j)$ pairs of quadrupole and angular momentum depicted in, 
e.g., Fig.~\ref{fig:plotrome} are physically meaningful.
The maximum rotation rate of a neutron star taking into account both the effects of general relativity and deformations has been found to be $\nu_{\rm s,max} = 1045 (M_0/M_\odot)^{1/2}(10\,{\rm km}/R)^{3/2}$ Hz, largely independent on the EoS \citep[see][for details]{2004Sci...304..536L}. Corresponding to this maximum rotation rate, the angular momentum is $J_{\rm max} = 2\pi \nu_{\rm s,max} I \sim 6.56\times 10^{48} I_{45}$ g cm$^2$ s$^{-1}$, and $j_{\rm max}=G J_{\rm max}/(c M^2_0)\sim 0.74 I_{45}/(M_0/M_\odot)^2$, where $I_{45}$ is the moment of inertia of the NS in units of $10^{45}$ g cm$^2$. The fastest observed pulsar is PSR J1748-2246ad with a rotation frequency of 716 Hz \citep{2006Sci...311.1901H}, which constrains the mass of the NS to $M_0 \geq 0.47 (R/10\,{\rm km})^3 M_\odot$, and $j\sim 0.51 I_{45}/(M_0/M_\odot)^2$, which becomes $j\sim 0.26 I_{45}$ for a canonical NS of $M_0=1.4 M_\odot$.

\subsection{Keplerian Frequency}

Now we turn into the frequencies analysis. For stationary axially symmetric spacetimes, the frequency of
Keplerian orbits is given by \citep[see e.g.][]{1995PhRvD..52.5707R}
\begin{equation}
\label{equ:DefOmegaK}
\Omega_{\rm K} = \frac{-g_{t\phi,\rho} \pm \sqrt{g_{t\phi,\rho}^2 - g_{\phi\phi,\rho} g_{tt,\rho}}}{g_{\phi\phi,\rho}},
\end{equation}
where a colon stands for partial derivative with respect to the indicated coordinate and `+' and 
`-' stands for corotating and counter-rotating orbits, respectively.

For the case of static spacetimes, i.e. for $\omega=0$ and therefore $g_{t\phi} = 0$, 
$\Omega_{\rm K} = \pm \sqrt{-g_{\phi\phi,\rho} g_{tt,\rho}}/g_{\phi\phi,\rho}$ and 
the energy $E$ and angular momentum $L$ per mass $\mu$ of the test particle can be expressed 
in terms of the metric tensor components \citep[see e.g.][]{1995PhRvD..52.5707R},
\begin{align}
\frac{E}{\mu} &= \frac{-g_{tt}}
{\sqrt{-g_{tt} -g_{\phi\phi} \Omega_{\rm K}^2}},
\qquad
\frac{L}{\mu} =\frac{g_{\phi\phi} \Omega_{\rm K}}
{\sqrt{-g_{tt} -g_{\phi\phi} \Omega_{\rm K}^2}}.
\end{align}
From here, it is clear that taking the negative branch of the root for $\Omega_{\rm K}$ in Eq.~(\ref{equ:DefOmegaK}) is equivalent to studying a particle with opposite angular momentum, i.e. $L_{\rm count-rot} = - L_{\rm co-rot}$. Thus, in the static case the magnitude of the energy and angular momentum are invariant under the change $\Omega_{\rm K} \rightarrow -\Omega_{\rm K}$. 

Now we consider the case of stationary space times, $\omega \ne 0$. The energy $E$ and angular momentum $L$ per mass $\mu$ are, in this case, given by \citep[see e.g.][]{1995PhRvD..52.5707R}
\begin{align}
\label{equ:E}
\frac{E}{\mu} &= \frac{-g_{tt} -g_{t\phi} \Omega_{\rm K}}
{\sqrt{-g_{tt} -2g_{t\phi} \Omega_{\rm K}-g_{\phi\phi} \Omega_{\rm K}^2}},
\\
\label{equ:L}
\frac{L}{\mu} &=\frac{g_{t\phi} + g_{\phi\phi} \Omega_{\rm K}}
{\sqrt{-g_{tt} -2 g_{t\phi} \Omega_{\rm K}-g_{\phi\phi} \Omega_{\rm K}^2}}.
\end{align}

The counter-rotating condition expressed by the negative branch of Eq.~(\ref{equ:DefOmegaK}), can be generated by the change $g_{t\phi} \rightarrow -g_{t\phi}$, which seems to be a more physical and transparent condition. In contrast to the static case, the counter-rotating orbit has now different energy and different magnitude of the angular momentum due the presence of the dragging of inertial frames, characterized by the metric component $g_{t\phi}$ (cf. Eq.~(\ref{equ:wLT}) below). In a nutshell, the dynamics of counter-rotating orbits of a test-particle can be derived, starting from the positive branch of Eq.~(\ref{equ:DefOmegaK}), by considering a spacetime with $g_{t\phi} \rightarrow -g_{t\phi}$.

For the vacuum case, a similar analysis as the one deve\-lo\-ped by \citet{herrera2006}, clearly shows that the change in the global sign of $g_{t\phi}$ is achieved by changing not only the angular momentum of the star, $J\rightarrow -J$, but \emph{all the rotational multipolar moments}. For the Kerr metric this change is obtained by changing the sign of the parameter $a$ (see Appendix~ \ref{app:Kerrmetricfuncs}) while in the PRS solution we need additionally change the sign of the parameter $s$ associated to \emph{differential rotation}, i.e., by changing $a\rightarrow -a$ and $s \rightarrow -s$\footnote{For the vacuum case, in the solution by \cite{2000PhRvD..62d4048M}, the sign change of $g_{t\phi}$ is obtained after performing simultaneously the replacements $a\rightarrow -a$  
and $b \rightarrow -b$.}. 

\begin{figure}
\centering
\includegraphics[width=\columnwidth,clip]{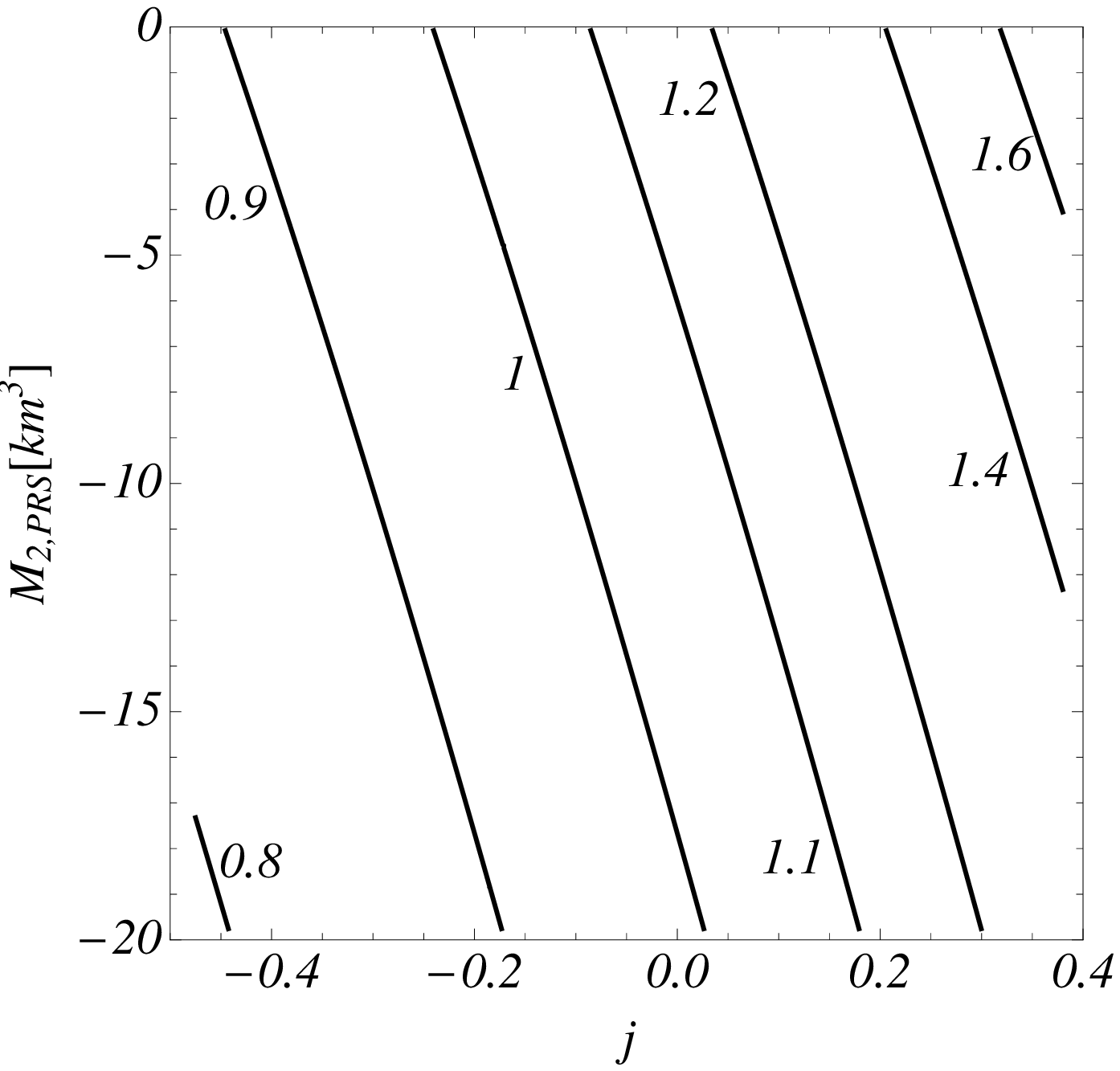}\\
\includegraphics[width=\columnwidth,clip]{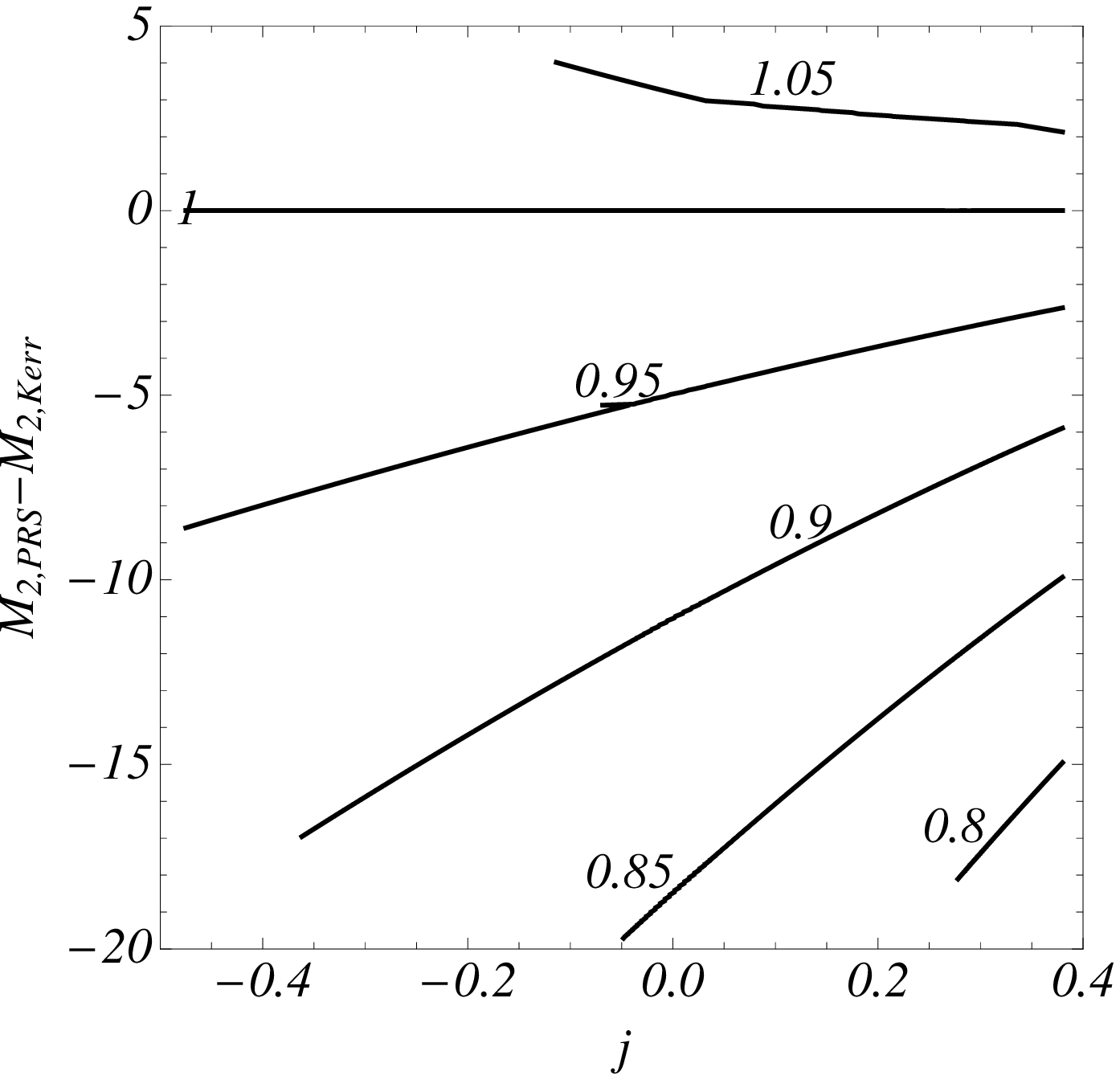}
\caption{Upper panel: Contours of constant $\nu_{\rm K}$ (in kHz) as a function of the the dimensionless angular momentum parameter $j=J/M^2_0$ and quadruple moment $M_2$ for the PRS solution, at the ISCO radius, for a compact object with mass $M_0=m=1.88 M_{\odot}=2.78$ km. Lower panel: Contours of constant ratio $\nu_{\rm K,PRS}/\nu_{\rm K,Kerr}$ as a function of $j$ and the difference $M_{2,\rm PRS}-M_{2,\rm Kerr}$, at the ISCO radius.}
\label{fig:plotomegaK}
\end{figure}

Once we have clarified this important issue about the co-rotating and counter-rotating orbits, we proceed to analyze the functional dependence of the Keplerian frequency on the multipole moments. In the upper panel of Fig.~\ref{fig:plotomegaK} we plotted contours of constant Keplerian frequency for the PRS solution, $\nu_{\rm K,PRS}=\Omega_{\rm K,PRS}/(2\pi)$, as a function of the dimensionless angular momentum parameter $j$ and the quadrupole moment $M_{2,\rm PRS}$, at the ISCO radius. It can be seen that the influence of the quadrupole moment is non-negligible, as evidenced from the departure of the contour lines from vertical lines. The Keplerian frequency grows for increasing $J$ and $M_2$. In the lower panel, we plotted contours of constant ratio $\nu_{\rm K,PRS}/\nu_{\rm K,Kerr}$ as a function of $j$ and the difference between the quadrupole moment of the PRS solution, $M_{2,\rm PRS}$, and the Kerr quadrupole, $M_{2,\rm Kerr}$. 

It is appropriate to recall here that because the Keplerian as well as the other frequencies calculated below are evaluated using formulas in the coordinate frame, see for instance Eq.~(\ref{equ:DefOmegaK}), they must be evaluated at coordinate radii $\rho$ and not at physical radii given by $\sqrt{g_{\phi\phi}}$. In the specific case of the ISCO the frequencies are evaluated at the radius that simultaneously solves the equations $V=0$, $\mathrm{d}V/\mathrm{d}\rho = 0$, and $\mathrm{d}^2V/\mathrm{d}\rho^2 = 0$, where $V$ is the effective potential (\ref{equ:EffPot}). 

\subsection{Oscillation and Precession Frequencies}

The radial and vertical oscillation (or epicyclic) frequencies are the frequencies at which the periastron and orbital plane of a circular orbit oscillates if we apply slightly radial and vertical perturbations to it, respectively. According to \cite{1995PhRvD..52.5707R}, in stationary axially symmetric vacuum spacetimes described by the Weyl-Papapetrou metric (\ref{Papapetrou}), the radial and vertical epicyclic frequencies can be obtained as
\begin{align}
\label{omegas}
\nu^{\rm OS}_{\alpha} &= \frac{1}{2\pi}\left\{ -\frac{g^{\alpha\alpha}}{2} \left[
(g_{tt} + g_{t \phi} \Omega)^2
    \left(\frac{g_{\phi \phi}}{\rho^2} \right)_{,\alpha \alpha} 
\right. \right.\nonumber \\
&-  2 (g_{tt} + g_{t \phi} \Omega)(g_{t \phi} + g_{\phi \phi} \Omega)
    \left(\frac{g_{t \phi}}{\rho^2} \right)_{,\alpha \alpha}
\nonumber \\
&+  \left. \left. (g_{t \phi} + g_{\phi \phi} \Omega)^2
    \left(\frac{g_{t t}}{\rho^2} \right)_{,\alpha \alpha} {} \right] \right\}^{1/2},
\end{align}
and the corresponding periastron ($\nu^{\rm P}_{\rho}$) and nodal ($\nu^{\rm P}_{z}$) precession frequencies as
\begin{equation}
\label{omegasP}
\nu^{\rm P}_{\alpha} = \nu_{\rm K}-\nu^{\rm OS}_{\alpha}\, .
\end{equation}
where $\alpha=\{\rho,z\}$, respectively, and $\nu_{\rm K}=\Omega_{\rm K}/(2 \pi)$ is the Keplerian orbital frequency with $\Omega_{\rm K}$ given by Eq.~(\ref{equ:DefOmegaK}).

\begin{figure}
\centering
\includegraphics[width=\columnwidth,clip]{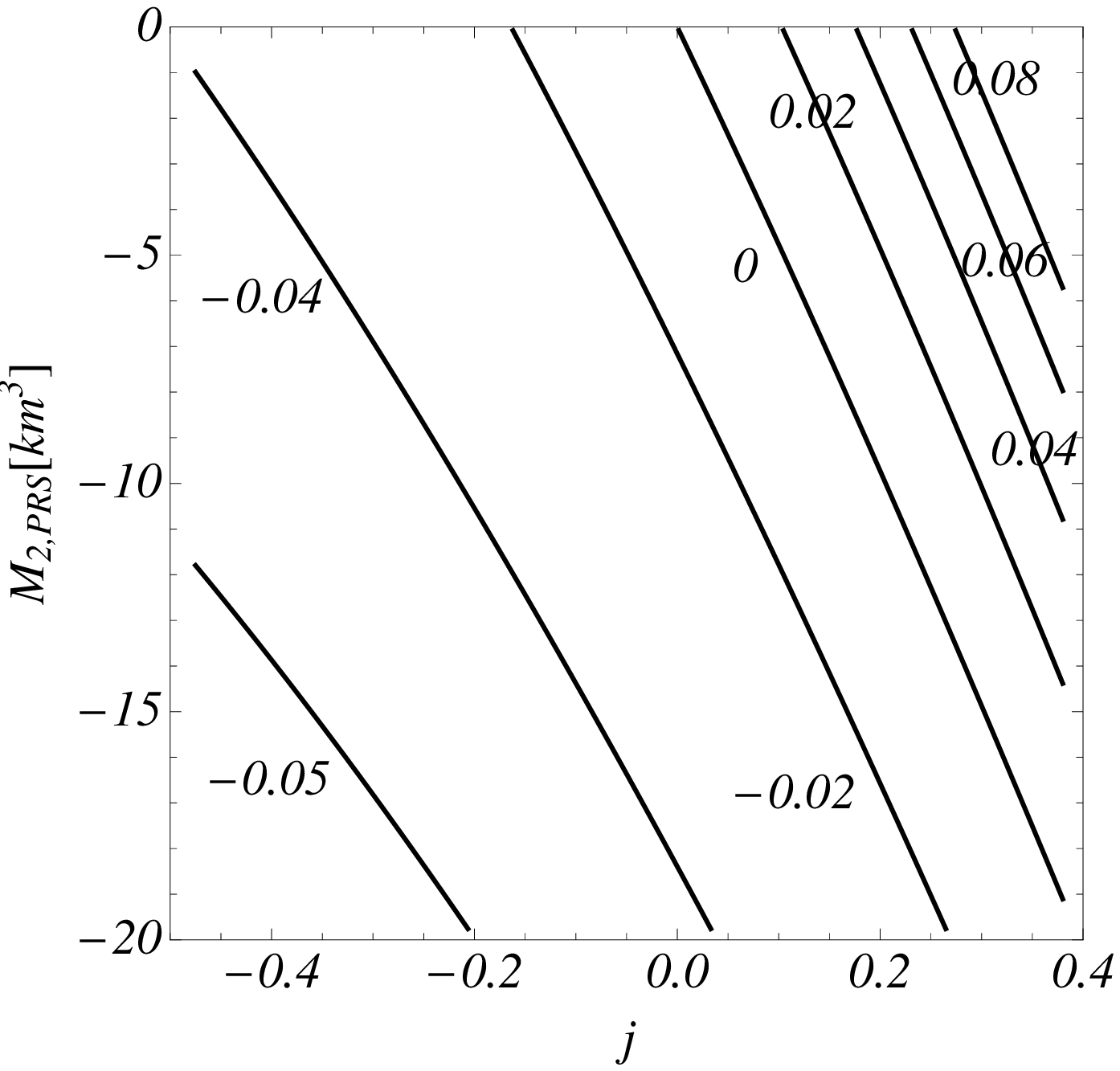}\\
\includegraphics[width=\columnwidth,clip]{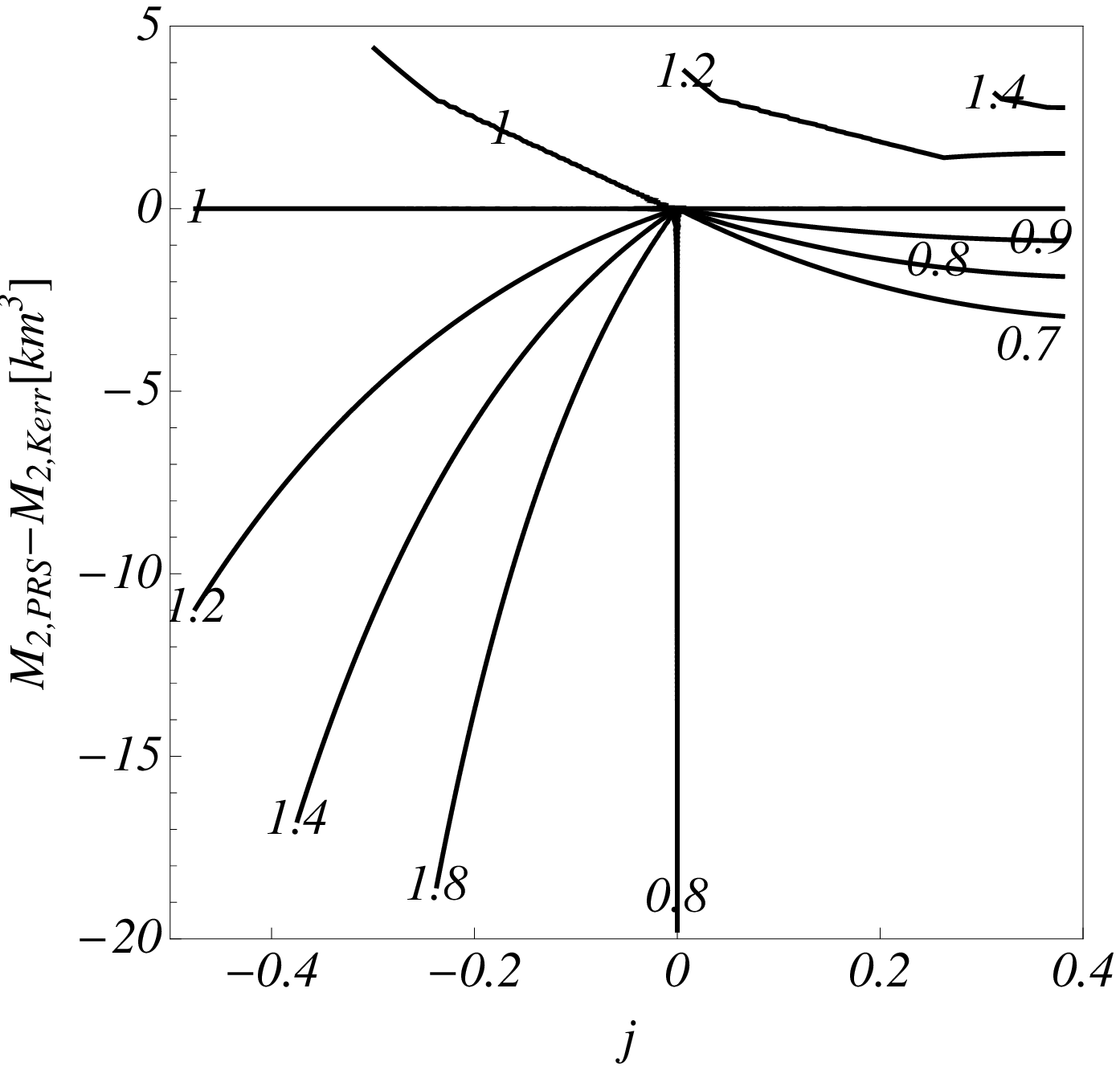}
\caption{Upper panel: $ \nu^{\rm P}_{z}$ (in kHz) as a function of the the dimensionless angular momentum parameter $j=J/M^2_0$ and quadruple moment $M_2$ for the PRS solution, at the ISCO radius, for a compact object with mass $M_0=m=1.88 M_{\odot}=2.78$ km. Lower panel: Contours of constant ratio $\nu^{\rm P}_{z,\rm PRS}/\nu^{\rm P}_{z,\rm Kerr}$ as a function of $j$ and the difference $M_{2,\rm PRS}-M_{2, \rm Kerr}$, at the ISCO radius.}
\label{fig:plotomegaz}
\end{figure}

In the upper panel of Fig.~\ref{fig:plotomegaz}, we plotted contours of constant nodal precession frequency $\nu^{\rm P}_z$ at the ISCO radius as a function of $j=J/M^2_0$ and $M_2$ for the PRS solution, at the ISCO radius. We can see now that the influence of the quadrupole moment is quite important. The nodal precession frequency increases for increasing $J$ and decreasing $M_2$, at fixed $M_0$. In the lower panel we plotted contours of constant ratio $\nu^{\rm P}_{z,\rm PRS}/\nu^{\rm P}_{z,\rm Kerr}$, at the ISCO radius, as a function of $j$ and the difference $M_{2,\rm PRS} - M_{2,\rm Kerr}$, in order to evidentiate deviations from the Kerr solution. The radial oscillation frequency $\nu^{\rm OS}_\rho$ \emph{vanishes at the ISCO radius} and therefore at such location the radial precession frequency equals the Keplerian frequency, whose contours have been plotted in Fig.~\ref{fig:plotomegaK}. 

\begin{figure}
\centering
\includegraphics[width=\columnwidth,clip]{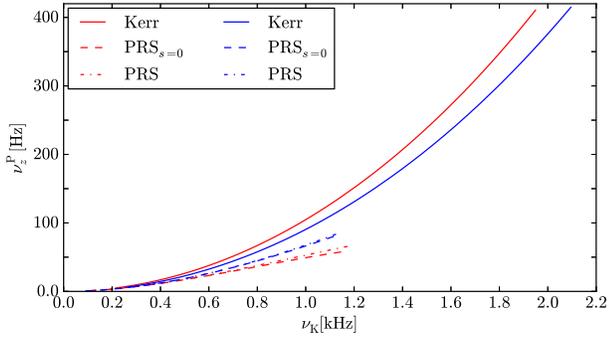}
\caption{Nodal precession frequency $\nu^{\rm P}_z$ versus Keplerian frequency $\nu_{\rm K}$ given by the Kerr and PRS analytic solutions. The lowest multipole moments have been fixed from the rotating NS models 20 (red curves) and 26 (blue curves) of the Table VI of \cite{PA12} for the EoS L. Model 20: $M_0=4.167$ km ($2.82 M_\odot$), $j=J/M^2_0 = 0.70$, $M_2 = -79.8$ km$^3$ ($-1.08\times 10^{45}$ g cm$^2$) and $S_3=-401.0$ km$^4$. Model 26: $M_0=4.36$ km ($2.95 M_\odot$), $j=J/M^2_0 = 0.56$, $M_2 = -45.2$ km$^3$ ($-6.10\times 10^{44}$ g cm$^2$) and $S_3=-170.0$ km$^4$.}
\label{fig:UpDownFreq1}
\end{figure}

\begin{figure}
\centering
\includegraphics[width=\columnwidth,clip]{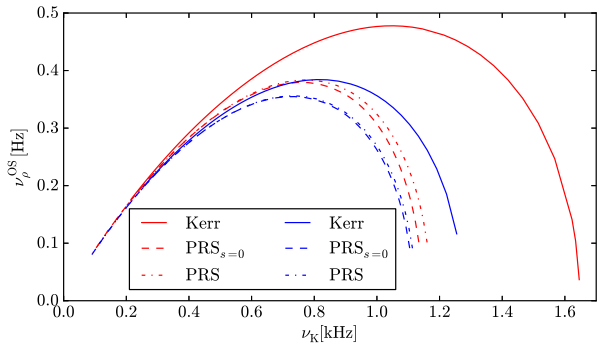}
\caption{Radial oscillation frequency $\nu^{\rm OS}_\rho$ versus Keplerian frequency $\nu_{\rm K}$ given by the Kerr and PRS analytic solutions. The lowest multipole moments has been fixed from the rotating NS models 20 (red curves) and 26 (blue curves) of the Table VI of \cite{PA12} for the EoS L. Model 20: $M_0=4.167$ km ($2.82 M_\odot$), $j=J/M^2_0 = 0.70$, $M_2 = -79.8$ km$^3$ ($-1.08\times 10^{45}$ g cm$^2$) and $S_3=-401.0$ km$^4$. Model 26: $M_0=4.36$ km ($2.95 M_\odot$), $j=J/M^2_0 = 0.56$, $M_2 = -45.2$ km$^3$ ($-6.10\times 10^{44}$ g cm$^2$) and $S_3=-170.0$ km$^4$.}
\label{fig:UpDownFreq2}
\end{figure}

In Figs.~\ref{fig:UpDownFreq1} and \ref{fig:UpDownFreq2} we plotted the nodal precession frequency $\nu^{\rm P}_z$ and the radial oscillation frequency $\nu^{\rm OS}_\rho$ as a function of the Keplerian frequency $\nu_{\rm K}$, respectively, for both the Kerr and PRS solutions. As an example, we have shown the results for rotating NS models 20 and 26 of Table VI of \cite{PA12}, for the EoS L. The lowest multipole moments of the PRS solution $M_0$, $J$, $M_2$, and $S_3$ have been fixed to the numerical values obtained by \cite{PA12}. In the case of the Kerr solution, only $M_0$ and $J$ can be fixed, while $M_2$, and $S_3$ have values induced by the lower moments $M_0$ and $J$. For the PRS solution with $s=0$, $M_0$, $J$ and $M_2$ can be fixed while $S_3$ cannot be fixed and depends on the lower moments. The results for the PRS analytic model obtained by fixing $M_0$, $J$, $M_2$, as well as $S_3$ are also shown. 

The deviations of the quadrupole and current octupole moments given by the Kerr solution from the numerical values of \cite{PA12} can be used to show the low accuracy of the Kerr solution to describe fast rotating NSs. The accuracy of the PRS solution in describing the ISCO radii of these two models has been shown in Table \ref{tab:ISCO} of Section \ref{sec:3}. 

In Figs.~\ref{fig:UpDownFreq1} and \ref{fig:UpDownFreq2} we can see the differences of the $\nu^{\rm P}_z$--$\nu_{\rm K}$ and $\nu^{\rm OS}_\rho$--$\nu_{\rm K}$ relations between the Kerr and PRS solutions for \emph{realistic NS models}. 
The deviations of the Kerr solution, especially at fast rotation rates, are evident because of the influence of the deformation (quadrupole $M_2$) of the star as well as, although in less proportion, of the octupole current $S_3$. 
In general, we observe that the larger the angular momentum, the poorer the performance of the predictions of Kerr solution. 

We have also shown in Figs.~\ref{fig:UpDownFreq1}--\ref{fig:UpDownFreq2} the influence of the current octupole $S_3$ in the determination of the precession and oscillation frequencies. We found that the effect of $S_3$ is only appreciable for the fastest models. The minor influence, in this case, of the current octupole $S_3$ is expected from the small values of the parameter $s$ needed to fit the numerical values of \cite{PA12}. Clearly, larger values of the parameter $s$ needed to fit realistic values of $S_3$ will enhance as well deviations from the Kerr spacetime. 

\textcolor{black}{The effects of a multipolar structure that deviates from the one of the Kerr geometry on the various quantities analyzed here} are relevant for instance in the RPM of the QPOs observed in LMXBs (see e.g.~\cite{1998ApJ...492L..59S,1999ApJ...513..827M,1999ApJ...524L..63S,1999PhRvL..82...17S} and Section \ref{sec:RPM}, for details).

\subsection{Dragging of Inertial Frames}

It is known that a prediction of general relativity is that a rotating object makes a zero angular momentum test particle to orbit around it, namely it \emph{drags} the particle into the direction of its rotation angular velocity; such an effect is called dragging of inertial frames or Lense-Thirring effect. Consequently, oblique particle orbit planes with respect to the source equatorial plane will precess around the rotation axis of the object. In stationary axially symmetric spacetimes described by the metric (\ref{Papapetrou}) the frame dragging precession frequency is given by \citep[see e.g.][]{1995PhRvD..52.5707R}
\begin{equation}
\label{equ:wLT}
\nu_{\rm LT} = - \frac{1}{2\pi} \frac{g_{t\phi}}{g_{\phi \phi}}\, .
\end{equation}

Many efforts have been done to test the predictions of general relativity around the Earth such as the analysis of the periastron precession of the orbits of the LAser GEOdynamics Satellites, LAGEOS and LAGEOS II, \citep[see e.g.][]{2010PhRvL.105w1103L} and the relativistic precession of the gyroscopes on-board the Gravity Probe B satellite \citep[see][for details]{2011PhRvL.106v1101E}. The latter experiment measured a frame dragging effect within an accuracy of 19\% with respect to the prediction of general relativity. 

The smallness of this effect around the Earth makes such measurements quite difficult and has represented a multi year challenge for Astronomy. The frame dragging precession increases with the increasing of the angular momentum of the rotating object and therefore a major hypothetical arena for the searching of more appreciable Lense-Thirring precession is the spacetime around compact objects such as BHs and NSs. The much stronger gravitational field of these objects with respect to the Earth one allow them to attain much faster angular rotation rates and so larger angular momentum.

\cite{1998ApJ...492L..59S} showed how, in the weak field slow rotation regime, 
the vertical precession frequency $\nu^{\rm P}_z$ (orbital plane precession frequency) 
can be divided into one contribution due to the Lense-Thirring precession and another 
one due to the deformation (non-zero quadrupole moment) of the rotating object, both 
of them comparable from the quantitative point of view. 
These frequencies could be in principle related to the motion of the matter in the accretion 
disks around BHs and NSs and thus particularly applicable to LMXBs. For fast rotating 
NSs and BHs the frequency at which the orbital plane, and so the frame dragging precession 
frequency, can reach values of the order of tens of Hz (see e.g.~\cite{1998ApJ...492L..59S} 
and Figs.~\ref{fig:plotomegaz} and \ref{fig:UpDownFreq1}). 

Thus, it is clear that an observational confirmation of the relativistic precession of matter 
around either a NS or a BH will lead to an outstanding test of the general relativity in the strong 
field regime and, at the same time, an indirect check of the large effects of the frame dragging 
in the exterior spacetime of compact objects \citep[see e.g.][for details]{1999ApJ...513..827M}. 

Although making independent measurements of the frame dragging effect around BHs and 
NSs is a very complicate task, it is important to know the numerical values of the precession 
frequency due to the frame dragging with respect to other relativistic precession effects, e.g. 
geodetic precession. In addition, it is important to know the sensitivity of the precession 
frequency to the object parameters such as mass, angular momentum, quadrupole, and octupole 
moment.

\begin{figure}
\centering
\includegraphics[width=\columnwidth,clip]{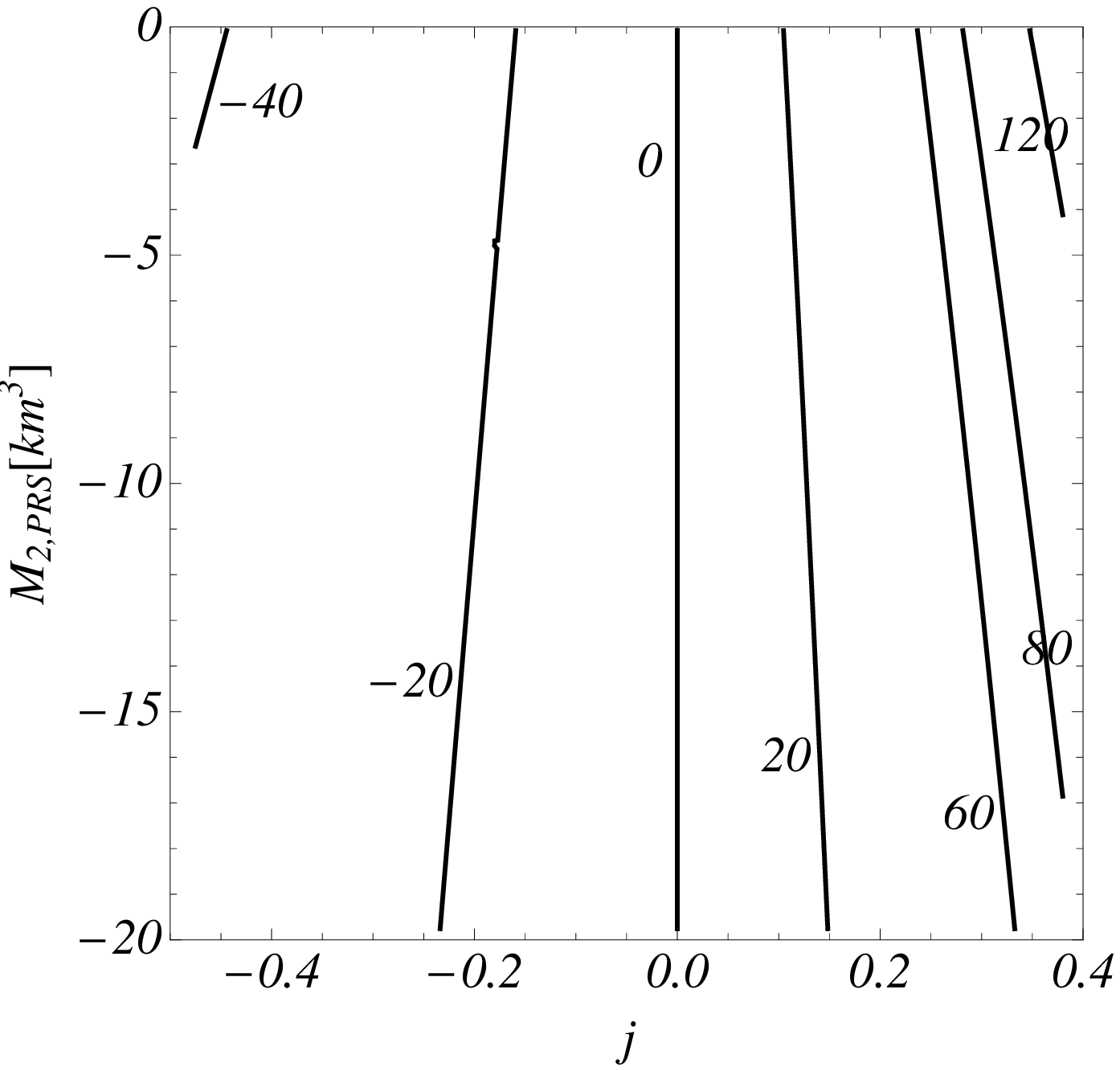}\\
\includegraphics[width=\columnwidth,clip]{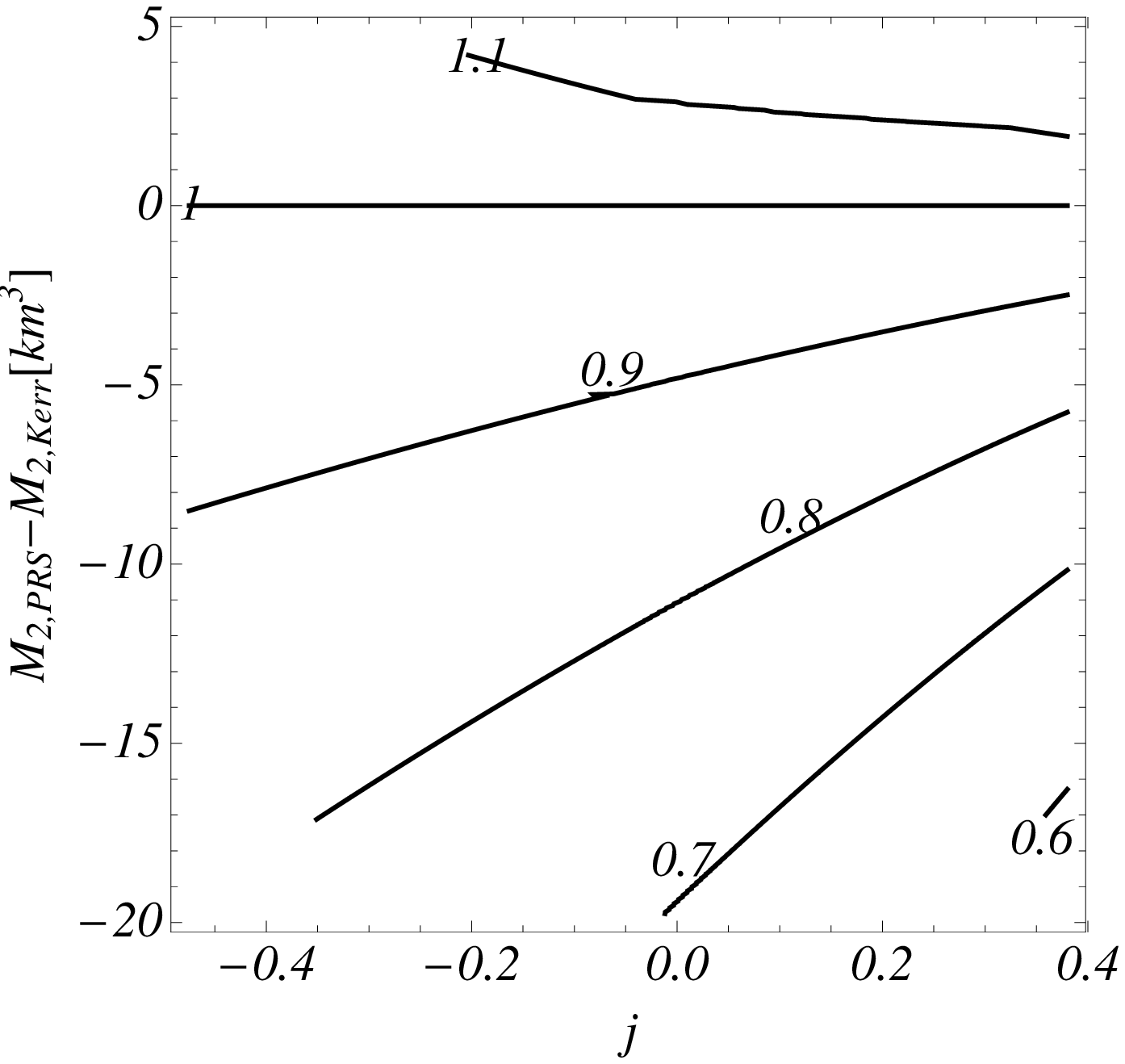}
\caption{Upper panel: Contours of constant $\nu_{\rm LT}$ (in Hz) as a function of the the angular momentum per unit mass $j=J/M^2_0$ and the quadruple moment $M_2$ for the PRS solution, at the ISCO radius, for a compact object with mass $M_0=m=1.88 M_{\odot}=2.78$ km. Lower panel: Contours of constant ratio $\nu_{\rm LT,PRS}/\nu_{\rm LT,Kerr}$ as a function of $j0$ and the difference $M_{2,\rm PRS}-M_{2,\rm Kerr}$, at the ISCO radius.}
\label{fig:plotomegaLT}
\end{figure}

In the upper panel of Fig.~\ref{fig:plotomegaLT} we plotted contours of constant frame dragging frequency $\nu_{\rm LT}$ for the PRS solution, at the ISCO radius, as a function of the the angular momentum per unit mass $J/M_0$ and the quadruple moment $M_2$, for a compact object mass $M_0=m=1.88 M_{\odot}$. Correspondingly, in the lower panel of Fig.~\ref{fig:plotomegaLT}, we show the differences between the frame dragging precession frequency as predicted by the Kerr and PRS solutions, at the ISCO radius, as a function of $j=J/M^2_0$ and the difference between the quadrupole moments, $M_{2,\rm PRS}-M_{2,\rm Kerr}$.

The influence of the quadrupole moment in the determination of the frame dragging frequency is evident; 
the frequency $\nu_{\rm LT}$ given by a NS is generally smaller than the one given by a BH as can be 
seen from the value of the ratio $\nu_{\rm LT,PRS}/\nu_{\rm LT,Kerr} < 1$ obtained for configurations 
with a quadrupole moment that deviates with respect to the one given by the Kerr solution, namely for 
$M_{2,\rm PRS}-M_{2,\rm Kerr} = M_{2,\rm PRS}+m a^2 = m k \neq 0$, see Eq.~(\ref{multipolosP}). 

It is also worth mentioning that frame dragging precession can be affected as well by the presence 
of electromagnetic fields \citep[see]{herrera2006} and further research in this respect deserves the 
due attention.

\section{Accuracy of Ryan's Analytic Formulas}\label{sec:4}

Following a series expansion procedure in powers of $1/\rho$, \cite{1995PhRvD..52.5707R} found that the periastron (radial) and nodal (vertical) precession frequencies, $\nu^{P}_\rho$ and $\nu^{P}_z$ given by Eq.~(\ref{omegas}), can be written as a function of the Keplerian frequency $\nu_{\rm K}$ as
\begin{align}
\label{eq:Ryan1}
&\frac{\nu^{\rm P}_\rho}{\nu_{\rm K}} = 
3 \mathcal{V}^2 - 
4 \frac{S_1}{M_0^2}\mathcal{V}^3 + 
\left( \frac{9}{2}-\frac{3}{2}\frac{M_2}{M_0^3} \right)\mathcal{V}^4 - 
10\frac{S_1}{M_0^2}\mathcal{V}^5 \nonumber \\
&+ \left( \frac{27}{2}-2\frac{S^2_1}{M_0^4}-\frac{21}{2}\frac{M_2}{M_0^3} \right)\mathcal{V}^6 + 
\left(-48\frac{S_1}{M_0^2}-5\frac{S_1 M_2}{M_0^5} \right.\nonumber \\
&\left. + 9 \frac{S_3}{M_0^4} \right)\mathcal{V}^7 + 
\left(\frac{405}{8}+\frac{2243}{84}\frac{S^2_1}{M_0^4}-\frac{661}{14}\frac{M_2}{M_0^3} 
\right.\nonumber \\ 
&\left.-\frac{21}{8}\frac{M^2_2}{M_0^6}+\frac{15}{4}\frac{M_4}{M_0^5} \right)\mathcal{V}^8 
+ \ldots\, ,
\end{align}
and
\begin{align}
\label{eq:Ryan2}
&\frac{\nu^{\rm P}_z}{\nu_{\rm K}} = 
2 \frac{S_1}{M_0^2}\mathcal{V}^3 + 
\frac{3}{2} \frac{M_2}{M_0^3}\mathcal{V}^4 +
 \left( 7\frac{S^2_1}{M_0^4}+3\frac{M_2}{M_0^3} \right)\mathcal{V}^6 \nonumber \\
&+ \left( 11\frac{S_1 M_2}{M_0^5}-6\frac{S_3}{M_0^4} \right)\mathcal{V}^7 + 
\left(\frac{153}{28}\frac{S^2_1}{M_0^4}+\frac{153}{28}\frac{M_2}{M_0^3} \right.\nonumber 
\\ &\left.+\frac{39}{8}\frac{M^2_2}{M_0^6}-\frac{15}{4}\frac{M_4}{M_0^5} \right)\mathcal{V}^8
+ \ldots\, ,
\end{align}
where $\mathcal{V}=(2\pi M_0 \nu_{\rm K})^{1/3}$, $[M_0,M_2,M_4]$ are the lowest three 
mass moments and, $[S_1,S_3]$,  are the lowest two current moments. For the PRS solution in the vacuum 
case, $M_4 = m(a^4 -3a^2 +k^2 + 2 a s)$. 

The above formulas are approximate expressions of the periastron and nodal precession frequencies in the weak field (large distances from the source) and slow rotation regimes. We should therefore expect that they become less accurate at distances close to the central object, e.g. at the ISCO radius, and for fast rotating objects. However, such formulas are an important tool to understand the role of the lowest multipole moments on the values of the relativistic precession frequencies, such as the importance of the higher multipole moments at short distances and high frequencies as can be seen from Eqs.~(\ref{eq:Ryan1}--\ref{eq:Ryan2}). 

\begin{figure}
\centering
\includegraphics[width=\columnwidth,clip]{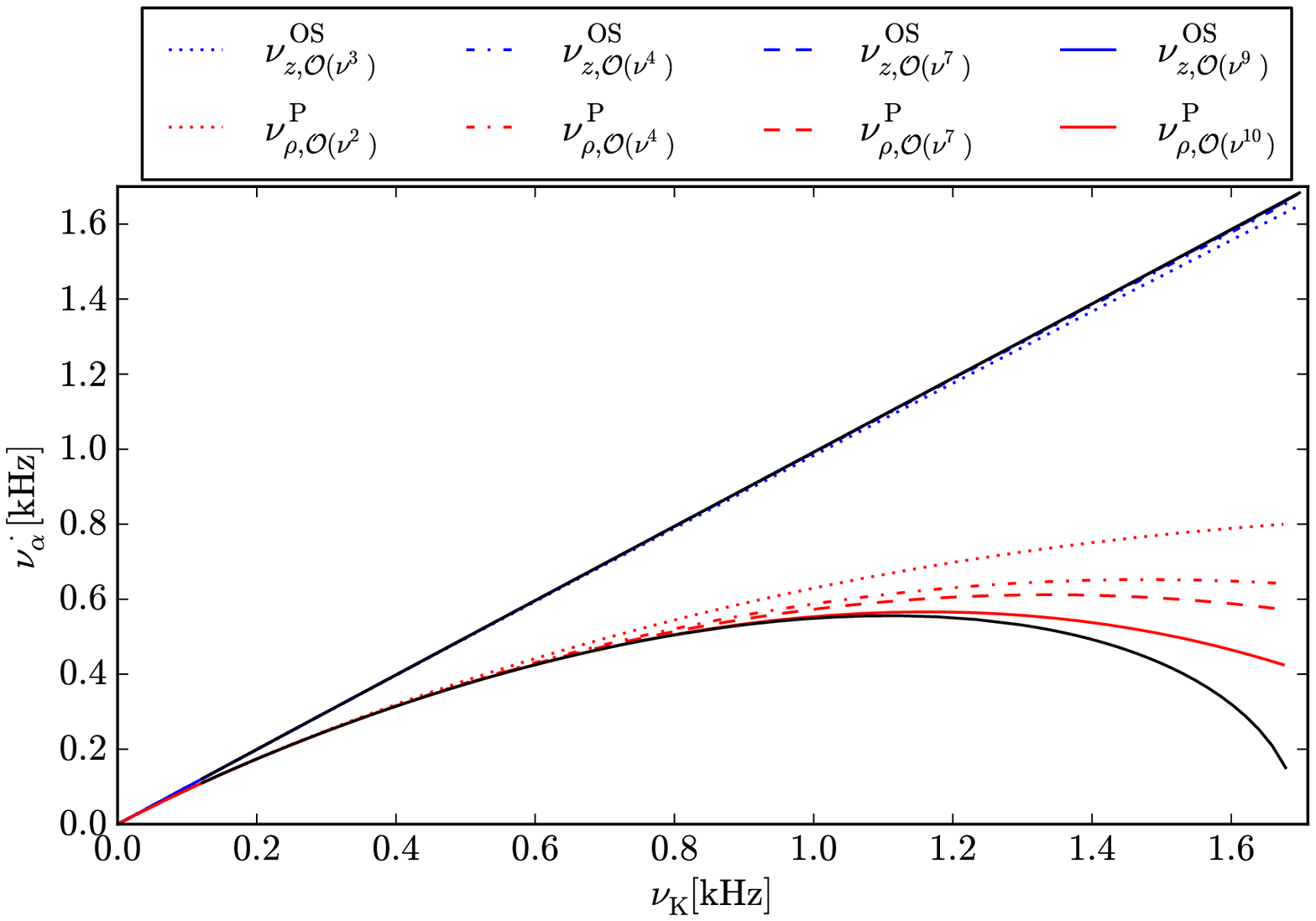}
\caption{Comparison of the $\nu^{\rm OS}_z$--$\nu_{\rm K}$ and $\nu^{\rm P}_\rho$--$\nu_{\rm K}$ relations given by the PRS solution and the approximate expressions (\ref{eq:Ryan1}--\ref{eq:Ryan2}) derived by \cite{1995PhRvD..52.5707R}. The lowest multipole moments $M_0$, $J$, $M_2$, and $S_3$ have been fixed to the values of the Model 2 of the Table VI of \cite{PA12}: $M_0= 2.071$ km ($1.402 M_\odot$), $j=0.194$, $M_2=-2.76$ km$^3$ ($3.73\times 10^{43}$ g cm$^2$), and $S_3=-2.28$ km$^4$.}
\label{fig:Ryan1}
\end{figure}

\begin{figure}
\centering
\includegraphics[width=\columnwidth,clip]{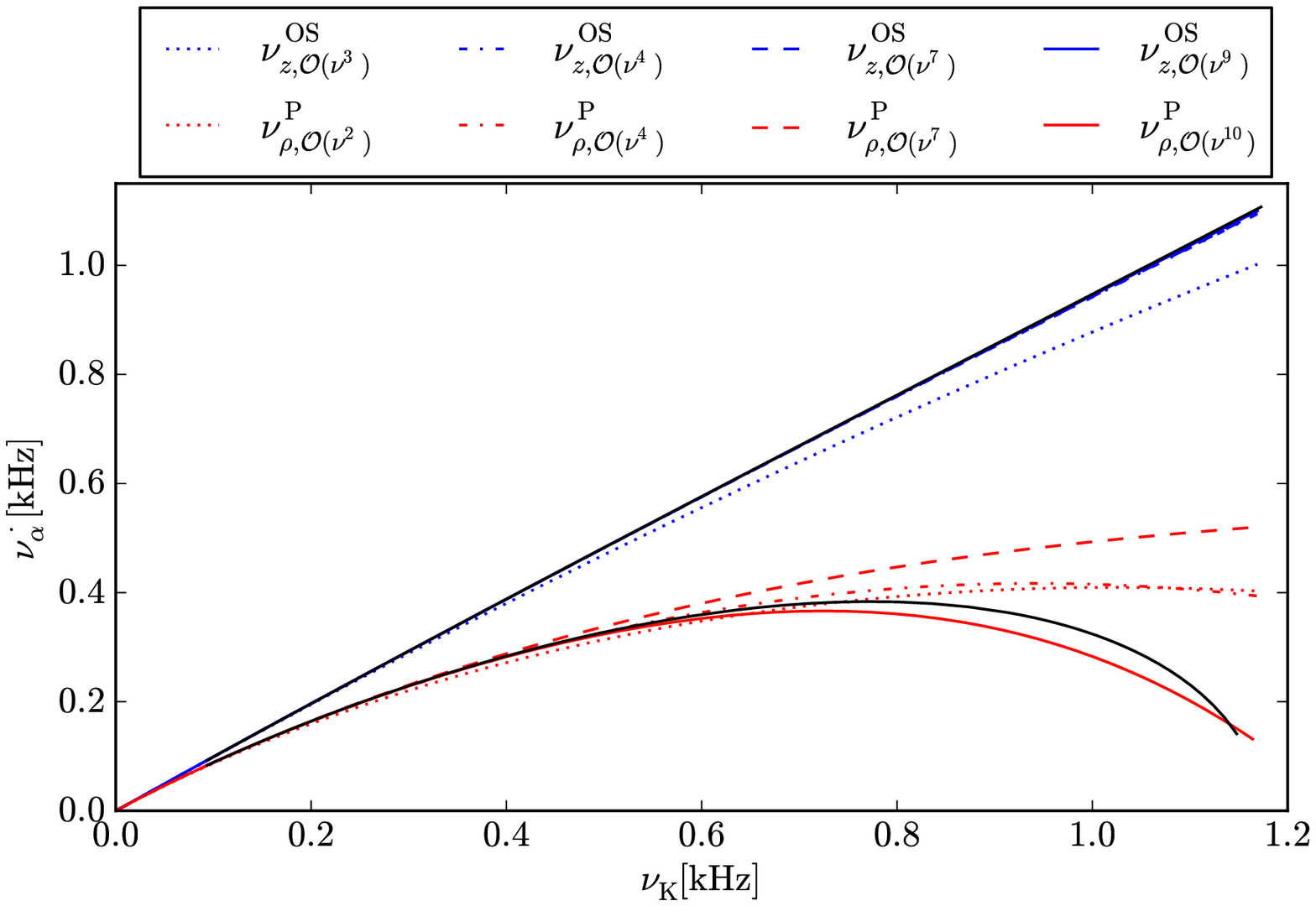}
\caption{Comparison of the $\nu^{\rm OS}_z$--$\nu_{\rm K}$ and $\nu^{\rm P}_\rho$--$\nu_{\rm K}$ relations as given by the PRS solution and the approximate expressions (\ref{eq:Ryan1}-\ref{eq:Ryan2}) derived by \cite{1995PhRvD..52.5707R}. The lowest multipole moments $M_0$, $J$, $M_2$, and $S_3$ have been fixed to the values of the Model 20 of the Table VI of \cite{PA12}: $M_0=4.167$ km ($2.82 M_\odot$), $j=J/M^2_0 = 0.70$, $M_2 = -79.8$ km$^3$ ($-1.08\times 10^{45}$ g cm$^2$) and $S_3=-401.0$ km$^4$}
\label{fig:Ryan2}
\end{figure}

At high frequencies, for instance of the order of kHz, deviations from the above scaling laws are appreciable. In Figs.~\ref{fig:Ryan1} and \ref{fig:Ryan2} we compare the radial precession and vertical oscillation frequencies, $\nu^{\rm P}_\rho$ and $\nu^{\rm OS}_z$, as a function of the Keplerian frequency $\nu_{\rm K}$, as given by the full expressions (\ref{omegas}) for the PRS solution and by the approximate formulas (\ref{eq:Ryan1}) and (\ref{eq:Ryan2}), respectively.\footnote{Because the scale of the $\nu^{\rm P}_\rho$ and $\nu^{\rm P}_z$ frequencies are very similar, we decided to plot in Fig.~\ref{fig:Ryan1} $\nu^{\rm P}_\rho$ and $\nu^{\rm OS}_z$ whose scales are different  allowing a more clear comparison with the PRS solution in a single figure.} The lowest multipole moments $M_0$, $J$, $M_2$, and $S_3$ of the PRS solution have been fixed to the values of two models of the Table VI of \cite{PA12}; Model 2 with $M_0= 2.071$ km ($1.402 M_\odot$), $j=0.194$, $M_2=-2.76$ km$^3$ ($3.73\times 10^{43}$ g cm$^2$), $S_3=-2.28$ km$^4$ and Model 20 with $M_0=4.167$ km ($2.82 M_\odot$), $j=J/M^2_0 = 0.70$, $M_2 = -79.8$ km$^3$ ($-1.08\times 10^{45}$ g cm$^2$) and $S_3=-401.0$ km$^4$.

In the $\nu^{\rm OS}_z$--$\nu_{\rm K}$ relation, the blue dotted curve depicts the contribution from 
the angular momentum (we plot the series (\ref{eq:Ryan2}) up to $\mathcal{V}^3$), for the blue 
dot-dashed curve we added the first contribution from the quadrupole moment $M_2$ (we cut the 
series at $\mathcal{V}^4$), for the dashed blue line we added the first contribution from the octupole 
mass-current (series expansion up to $\mathcal{V}^7$) and finally in the continuos blue line we 
consider contributions for higher multipole moments and stop the series at the order $\mathcal{V}^9$, 
not shown in Eqs.~(\ref{eq:Ryan2}). For this case, we can see that Ryan's expressions clearly tend, from the bottom, to the exact result (continuous black curve) obtained by using the PRS solution. 

For the analysis of the $\nu^{\rm P}_\rho$--$\nu_{\rm K}$ relation we followed the same procedure as 
described above. In this case, the Ryan's expressions tend from the top to the exact result, the continuous black curve, represented by the PRS solution. It is interesting to see that the introduction of the octupole moment (dashed red line) makes the approximation to deviate from the exact result, however by including more terms the accuracy is enhanced. As can be seen from Figs.~\ref{fig:Ryan1} and \ref{fig:Ryan2} the quantitative accuracy of the Ryan's 
approximate formulas in the periastron precession frequency $\nu^{\rm P}_\rho$ is less than the one 
obtained in the vertical oscillation frequency $\nu^{\rm OS}_z$.
 
The importance of the high-order multipole moments such as the quadrupole and the octupole moments is evident in the high-frequency regime. This is in line with the results shown in Figs.~\ref{fig:plotomegaK}--\ref{fig:plotomegaz} and in Figs.~\ref{fig:UpDownFreq1}--\ref{fig:UpDownFreq2}. We can see from Figs.~\ref{fig:Ryan1} and ~\ref{fig:Ryan2} that the Ryan's approximate formulas describe more accurately the Model 2 than the Model 20. The reason is that, as we mentioned above, we should expect a better accuracy of the series expansions from low to moderate moderate rotation rates and consequently the same occur for the quadrupole deformations. It is clear that there are appreciable differences both in rotation and deformation between the two selected models; we recall also that the rotation frequency of the star can be expressed as a function of the dimensionless $j$ parameter as $\nu_s = G j M^2_0/(2 \pi c I) = 1.4 (M/M\odot)^2/I_{45}$ kHz. 

It is noteworthy that we have checked that the Ryan's series expansions, Eqs.~(\ref{eq:Ryan1}) and (\ref{eq:Ryan2}), fit quite accurately the exact results if taken up to order $\mathcal{V}^{10}$. In particular the values of the vertical oscillation and precession frequencies are fit better than the corresponding radial ones. For the Model 2 the radial oscillation frequency is well fitted by the Ryan's expression up to Keplerian frequencies of order $\sim 1.2$ kHz while, for the Model 20, the approximate formulas break down at a lower value $\sim 0.7$ kHz. These results are of particular relevance because it makes possible the extraction of the object parameters (for instance the lowest multipoles up to $S_3$) by the fitting of the observed QPO frequencies in LMXBs, providing they are indeed related to the precession and oscillation frequencies of matter in the accretion disk (see Section \ref{sec:RPM}, for details) and for Keplerian motion not exceeding a few kHz of frequency.

\section{Accuracy of PRS solution}\label{sec:5}

We turn now to analyze the behavior of the Kerr and PRS solutions in predicting results for the Keplerian, frame dragging, and vertical oscillation frequencies, for realistic NSs. In particular, we compare their predictions with the frequencies calculated by \cite{1999ApJ...513..827M}. Since \cite{1999ApJ...513..827M} did not include the values of the octupole current moment $S_3$, here we set $s=0$ in Eq.~(\ref{Potenciales eje}) for the PRS solution. For the sake of comparison, we choose the results derived by \cite{1999ApJ...513..827M} for the EoS L, because for this EoS the highest rotating parameter $j$ and quadrupole moment $M_2$ were found. In addition, the stiffness of such an EoS allows the maximum mass of the NS to be larger than the highest observed NS mass, $M_0=1.97 \pm 0.04 M_\odot$, corresponding to the 317 Hz (3.15 milliseconds rotation period) pulsar J1614-2230 \citep[see][for details]{2010Natur.467.1081D}.

This regime of high $j$ and $M_2$ in realistic models is particularly interesting to test the deviations of the Kerr solution in the description of NS signatures as well as to explore the accuracy of the PRS solution. In Table~\ref{tab:FKLTZ}, we present the results for four different sets of the star spin frequency $\nu_s$, namely $\nu_s = 290$ Hz (M1 and M2), $\nu_s = 360$ Hz (M3 and M4), $\nu_s = 580$ Hz (M5 and M6) and $\nu_s = 720$ Hz (M7 and M8).
\begin{table*}[hbtp]
\centering
{\scriptsize
\begin{tabular}{cccccccccccccccc}
Model&$M_0/M_{\odot}$& $j$ & $M_2/Q_0$ & 
$r_{+}^{\rm MS}$[km] & $r_{+}^{\rm Kerr}$& $r_{+}^{\rm PRS}$& 
$\nu_{\rm K}^{\rm MS}$[kHz] & $\nu_{\rm K}^{\rm Kerr}$& $\nu_{\rm K}^{\rm PRS}$ & 
$\nu_{\rm LT}^{\rm MS}$[Hz] & $\nu_{\rm LT}^{\rm Kerr}$& $\nu_{\rm LT}^{\rm PRS}$ &
$\nu_{z}^{\rm P,MS}$[Hz] & $\nu_{z}^{\rm P,Kerr}$& $\nu_{z}^{\rm P,PRS}$\\
\tableline
M1& 1.88 & 0.19 & -5.3  & 15.4 & 14.90 & 15.42 & 1.31 & 1.363 & 1.304 & 39.7   & 42.248 & 38.476 & 18.6 & 39.697 & 22.040\\
M2& 2.71 & 0.14 & -3.0  & 22.2 & 22.16 & 22.23 & 0.90 & 0.906 & 0.902 & 19.6   & 19.676 & 19.493& 17.2 & 18.809 & 17.629\\
M3& 1.94 & 0.24 & -8.2  & 15.6 & 14.89 & 15.63 & 1.29 & 1.380 & 1.296 & 49.8   & 57.001 & 49.804& 17.5 & 52.608 & 26.197\\
M4& 2.71 & 0.18 & -4.8  & 21.8 & 21.62 & 21.74 & 0.93 & 0.937 & 0.931 & 26.1   & 27.245 & 26.833& 21.9 & 25.670 & 23.635\\ 
M5& 2.07 & 0.40 & -23.1 & 16.3 & 14.18 & 16.06 & 1.26 & 1.514 & 1.289 & 84.3   & 125.75 & 88.905& -10.5& 109.04 & 31.140\\
M6& 2.72 & 0.30 & -13.9 & 20.6 & 20.05 & 20.45 & 1.01 & 1.041 & 1.015 & 53.5   & 57.467 & 54.391& 35.8 & 51.861 & 42.854\\
M7& 2.17 & 0.51 & -39.4 & 17.0 & 13.58 & 16.53 & 1.22 & 1.637 & 1.269 & 106.7 & 201.52 & 116.25& -51.8 & 166.55 & 29.804\\
M8& 2.73 & 0.39 & -24.4 & 19.8 & 18.85 & 19.65 & 1.06 & 1.136 & 1.077 & 78.8   & 90.821 & 80.895& 38.4 & 79.079 & 57.187\\
\tableline
\end{tabular}
}
\caption{
ISCO radius $r_+$, Keplerian frequency $\nu_{\rm K}$, frame-dragging 
(Lense-Thirring) frequency $\nu_{\rm LT}$, and vertical precession frequency $\nu^{\rm P}_z$ of the co-rotating
orbits calculated numerically by \cite{1999ApJ...513..827M} (upper index MS) and comparison with the corresponding predicted values given by the Kerr (upper index Kerr) and the PRS$_{s=0}$ solution (upper index PRS). The quadrupole moment $M_2$ have been normalized for convenience to the value $Q_0 = 10^{43}$ g cm$^2$.
}\label{tab:FKLTZ}
\end{table*}

In Table~\ref{tab:FKLTZ}, we clearly observe that the results predicted by the PRS$_{s=0}$ solution for the Keplerian and frame-dragging frequencies are in excellent agreement with those calculated by \cite{1999ApJ...513..827M} for even highly massive, rotating and deformed models such as the model M7 with $M_0= 2.17 M_{\odot}$, $j=0.51$ and $M_2 = -39.4 Q_0$. We notice that \cite{1999ApJ...513..827M} reported some configurations with negative values of $\nu_z$ (see Table~\ref{tab:FKLTZ}). We advance the possibility that this is due to instabilities of the numerical code that occur when the ISCO radius is located very close or inside the surface of the object. Thus, the values of the frequencies given by the analytic solution in these cases are to be considered predictions to be tested for future numerical computations. This fact can be checked within the calculations of \cite{1999ApJ...513..827M} by exploring the properties of counter-rotating orbits which produce in general ISCO radii larger than the ones of the corotating ones. In Table~\ref{tab:FKLTZc}, we depicted the results in the counter-rotating case where we can notice an improvement of the accuracy of the PRS solution with respect to the co-rotating case. 

\begin{table*}[hbtp]
\centering
\begin{tabular}{cccccccccccccc}
Model& 
$r_{-}^{\rm MS}$[km] & $r_{-}^{\rm Kerr}$& $r_{-}^{\rm PRS}$& $\nu_{\rm K}^{\rm MS}$[kHz] & $\nu_{\rm K}^{\rm Kerr}$& $\nu_{\rm K}^{\rm PRS}$ & 
$\nu_{\rm LT}^{\rm MS}$[Hz] & $\nu_{\rm LT}^{\rm Kerr}$& $\nu_{\rm LT}^{\rm PRS}$ & $\nu_{z}^{\rm MS}$[Hz] & $\nu_{z}^{\rm Kerr}$& $\nu_{z}^{\rm PRS}$\\
\tableline
M1&  18.8 & 18.35 & 18.73 & 0.99 & 1.023 & 0.997 & 21.7  & 22.61 & 21.39 & 29.0 & 23.88 & 29.70\\
M2&  25.9 & 25.82 & 25.88 & 0.73 & 0.734 & 0.732 & 12.4  & 12.43 & 12.36 & 13.6 & 12.95 & 13.46\\
M3&  19.9 & 19.39 & 19.89 & 0.93 & 0.960 & 0.928 & 24.0  & 25.79 & 24.04 & 32.9 & 27.61 & 34.20\\
M4&  26.4 & 26.33 & 26.42 & 0.71 & 0.715 & 0.712 & 14.7  & 15.08 & 14.94 & 16.4 & 15.88 & 16.57\\ 
M5&  23.3 & 22.22 & 23.24 & 0.77 & 0.816 & 0.768 & 28.2 & 32.59 & 28.78 & 41.3 & 36.31 & 44.65\\
M6&  28.2 & 27.94 & 28.18 & 0.65 & 0.660 & 0.652 & 20.5 & 21.19 & 20.70 & 24.1 & 23.03 & 24.51\\
M7&  25.9 & 24.37 & 25.78 & 0.67 & 0.731 & 0.678 & 28.9 & 34.62 & 29.64 & 43.4 & 39.56 & 47.90\\
M8&  29.7 & 29.18 & 29.58 & 0.61 & 0.623 & 0.611 & 23.2 & 24.40 & 23.48 & 28.4 & 27.12 & 29.22\\
\tableline
\end{tabular}
\caption{
Same as in Table~\ref{tab:FKLTZ}, but for the counter-rotating case.
}\label{tab:FKLTZc}
\end{table*}

\textcolor{black}{
In this line, we consider worth performing} numerical computations of the precession and oscillation frequencies of particles around realistic NSs in a wider space of parameters and using up-to-date numerical techniques which will certainly help to establish and elucidate more clearly the accuracy of analytic models. It is also appropriate recalling the recent results of \cite{PA12} on the computation of the general relativistic multipole moments in axially symmetric spacetimes.  

\section{The Relativistic Precision Model}\label{sec:RPM}

The X-ray light curves of LMXBs show a variability from which a wide variety of QPOs have been measured, expanding from relatively low $\sim$ Hz frequencies  all the way up to high $\sim$ kHz frequencies \citep[see e.g.][for details]{1995LNP...454..321V}. In particular, such frequencies usually come in pairs (often called twin peaks), the lower and upper frequencies, $\nu_l$ and $\nu_h$ respectively. BHs and NSs with similar masses can show similar signatures and therefore the identification of the compact object in a LMXB is not a simple task. If the QPO phenomena observed in these systems are indeed due to relativistic motion of accretion disk matter, the knowledge of the specific behavior of the particle frequencies (e.g. rotation, oscillation, precession) in the exterior geometry of NSs and BHs becomes essential as a tool for the identification of the nature of the compact object harbored by a LMXB.

It is not the scope of this work to test a particular model for the QPO phenomenon in LMXBs but instead to show the influence of the high multipole moments on the orbital motion of test particles especially the role of the quadrupole moment which is of particular interest to differentiate a NS from a BH. There are in the literature several models that describe the QPOs in LMXBs through the frequencies of particles around the compact object, and for a recent review and comparison of the different models we refer to the recent work of \cite{2011ApJ...726...74L}. In order to show here the main features and differences between the Kerr and the PRS solutions we shall use the Relativistic Precession Model (RPM). 

The RPM model identifies the lower and higher (often called twin-peaks) kHz QPO frequencies, $\nu_l$ and $\nu_h$, with the periastron precession and Keplerian frequencies, namely $\nu_l = \nu^{\rm P}_{\rho}$ and $\nu_h = \nu_{\rm K}$, respectively. The so-called horizontal branch oscillations (HBOs), which belong to the low frequency QPOs observed in high luminosity Z-sources \citep[see e.g.][for details]{1995LNP...454..321V}, are related within the RPM model to the nodal precession frequency $\nu^{\rm P}_{z}$ of the same orbits \citep[see]{1999ApJ...513..827M}. We will use here in particular the realistic NS models of \cite{1999ApJ...513..827M} for the EoS L. 

One of the salient features of the RPM model is that in the case of the HBO frequencies, the relations inferred from the first term of the expansions (\ref{eq:Ryan1}) and (\ref{eq:Ryan2})
\begin{eqnarray}
\nu_{\rm K} &=& 3^{-3/5}(2\pi)^{-2/5}m^{-2/5} (\nu^{\rm P}_\rho)^{3/5}\, ,\\
\nu^{\rm P}_z &=& (2/3)^{6/5} \pi^{1/5} j\, m^{1/5} (\nu^{\rm P}_\rho)^{6/5}\, ,
\end{eqnarray}
which implies a nodal precession frequency proportional to the square of the Keplerian frequency has been observed in some sources, for instance in the LMXB 4U 1728--34 \citep[see][for details]{1998ApJ...506L..39F}. In addition, $6/5$ power law relating the nodal and periastron precession frequencies can explain \citep[see][]{1999ApJ...524L..63S} the correlation between two of the observed QPO frequencies found in the fluxes of NSs and BHs LMXBs \citep[see][for details]{1999ApJ...520..262P}. This fact provides, at the same time, a significant test of the Ryan's analytic expressions.

\begin{figure}
\centering
\includegraphics[width=\columnwidth,clip]{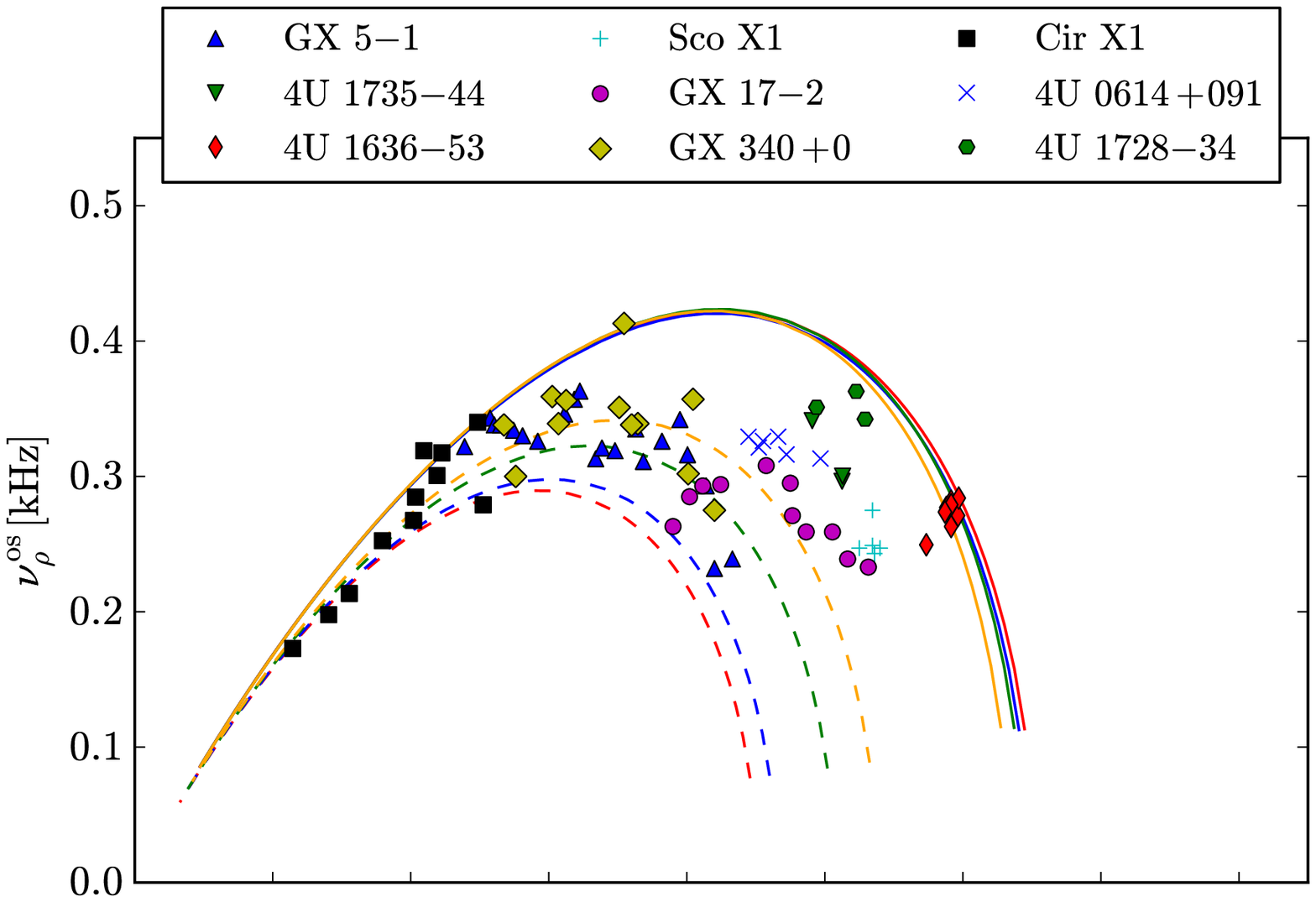}\\
\includegraphics[width=\columnwidth,clip]{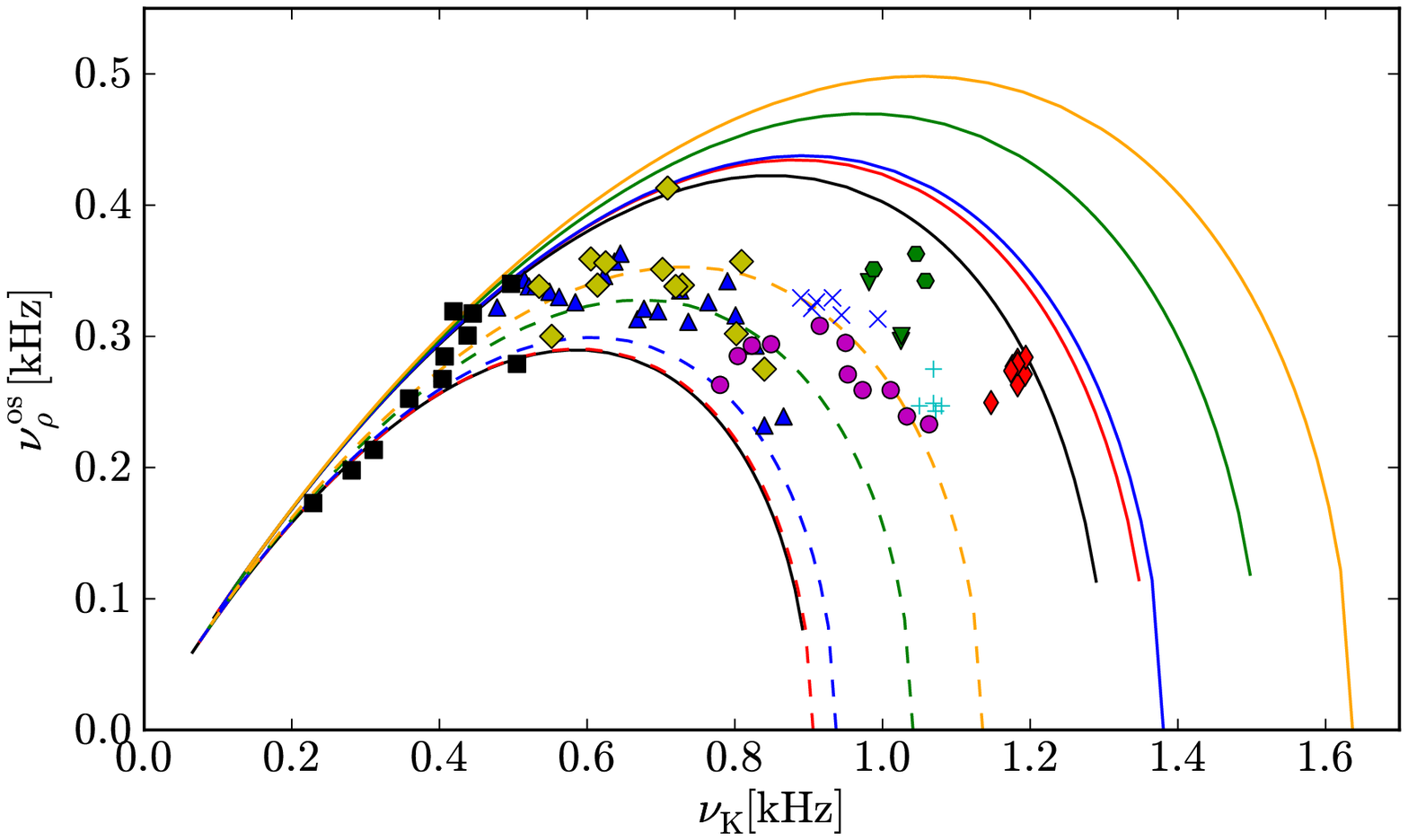}
\caption{Periastron oscillation frequency, $\nu^{\rm OS}_{\rho}$, as a function of the Keplerian frequency $\nu_{\rm K}$ for the NS realistic models in Table \ref{tab:FKLTZ}. We indicate the QPO frequencies observed in the sources GX 5--1, 4U 1735--44, 4U 1636--53, Sco X1, GX 17--2, GX 340+0, Cir X1, 4U 0614+091, and 4U 1728--34. The solid curves depict the results for the models M1 (solid) and M2 (dashed) with red lines, for the models M3 (solid) and M4 (dashed) with blue lines, for the models M5 (solid) and M6 (dashed) with green lines while orange lines stands for the results from models M7 (solid) and M8 (dashed). In the upper panel we present the results derived from the PRS$_{s=0}$ solution while in the lower panel we present the results for the Kerr solution. \textcolor{black}{In the lower panel we have added, to guide the eye, the inner red dashed and outer red solid curves of the upper panel using black lines.}}
\label{fig:plotomegarhoSMEOSL}
\end{figure}

It is interesting to analyze the level of predictability of the precession and oscillation frequencies on particular astrophysical sources. In Fig.~\ref{fig:plotomegarhoSMEOSL} we show the $\nu_l$--$\nu_h$ relation within the RPM model, namely $\nu^{\rm P}_{\rho}$ versus $\nu_{\rm K}$ for the models M1--M8 of Table \ref{tab:FKLTZ}. In the upper panel we show the results for the PRS solution while, in the lower panel, we present the results for the Kerr solution. We have indicated the QPO frequencies observed in the sources GX 5--1 \citep[see e.g.][]{1998ApJ...504L..35W,2002MNRAS.333..665J}, 4U 1735--44 \citep[see e.g.][]{1998ApJ...508L.155F}, 4U 1636--53 \citep[see e.g.][]{1997ApJ...479L.141W}, Sco X1 \citep[see e.g.][]{1996ApJ...469L...1V}, GX 17--2 \citep[see e.g.][]{2002ApJ...568..878H}, GX 340+0 \citep[see e.g.][]{2000ApJ...537..374J}, Cir X1 \citep[see e.g.][]{1996ApJ...469L...1V}, 4U 0614+091 \citep[see e.g.][]{1997ApJ...486L..47F}, and 4U 1728--34 \citep[see e.g.][]{1996ApJ...469L...9S}.
 
\textcolor{black}{Both the upper and lower panels of Fig.~\ref{fig:plotomegarhoSMEOSL} have been plotted using the same frequency scales in order to aid the identification of the differences between the Kerr and the PRS solutions. One can notice that all the solid curves in the Kerr solution (lower panel of Fig.~\ref{fig:plotomegarhoSMEOSL}) are outside the range of the observed QPO frequencies exemplified, while all dashed and solid curves of the PRS are inside the QPO range. It is then clear that making a fit of the observed QPO frequencies of the selected LMXBs of Fig.~\ref{fig:plotomegarhoSMEOSL} will necessarily require a different choice of parameters in the Kerr and PRS solutions.  Therefore, conclusions for instance on the NS parameters (e.g.~mass, angular momentum, quadrupole deformation) based on fitting QPOs using the Kerr geometry will deviate from the actual parameters \citep[see e.g.][for details]{1999ApJ...512..282L}, extractable more reliably from a more complex geometry, such as the PRS one, that allows a better estimate for instance of the quadrupole moment of a compact star.} 

\begin{figure}
\centering
\includegraphics[width=\columnwidth,clip]{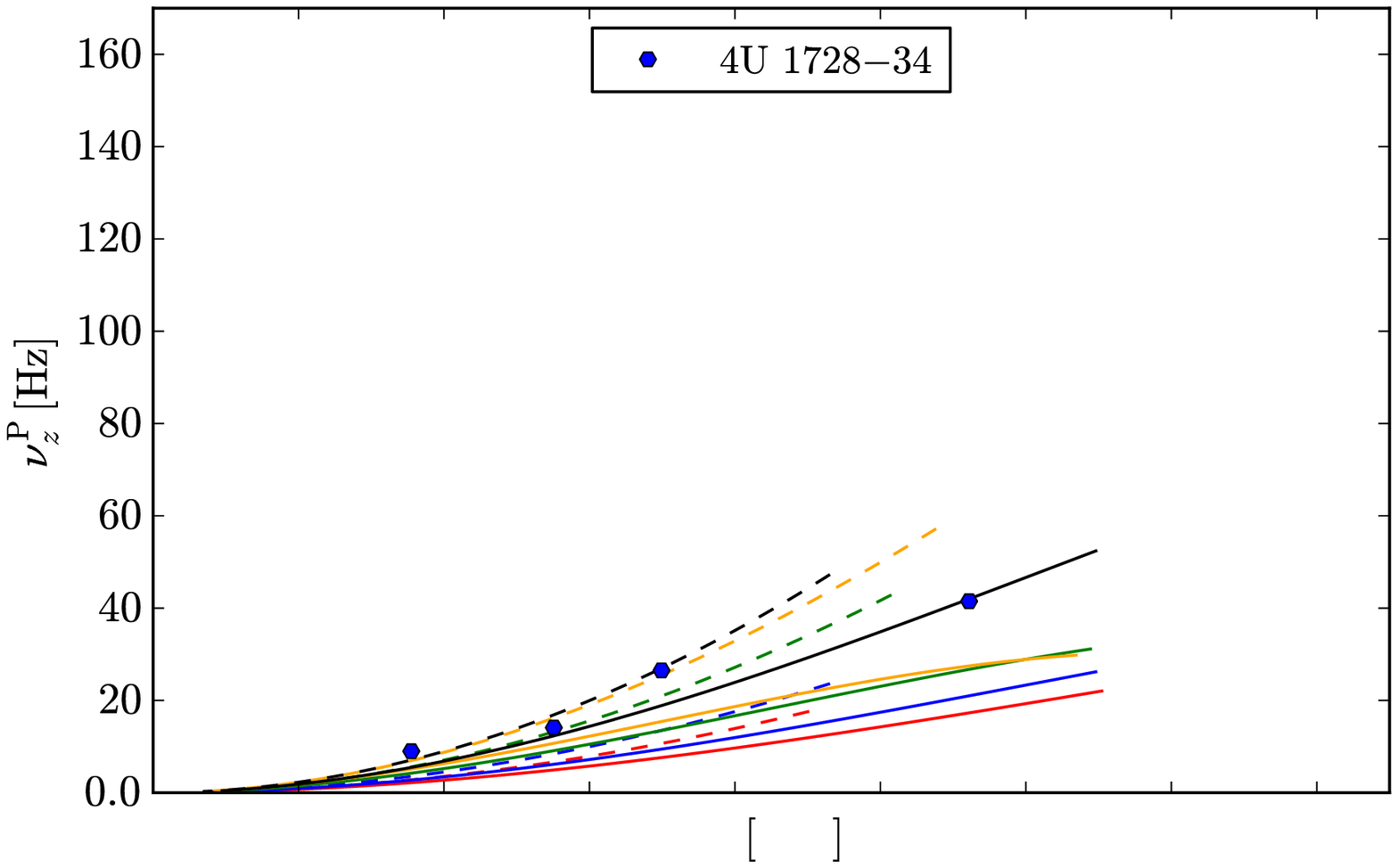}\\
\includegraphics[width=\columnwidth,clip]{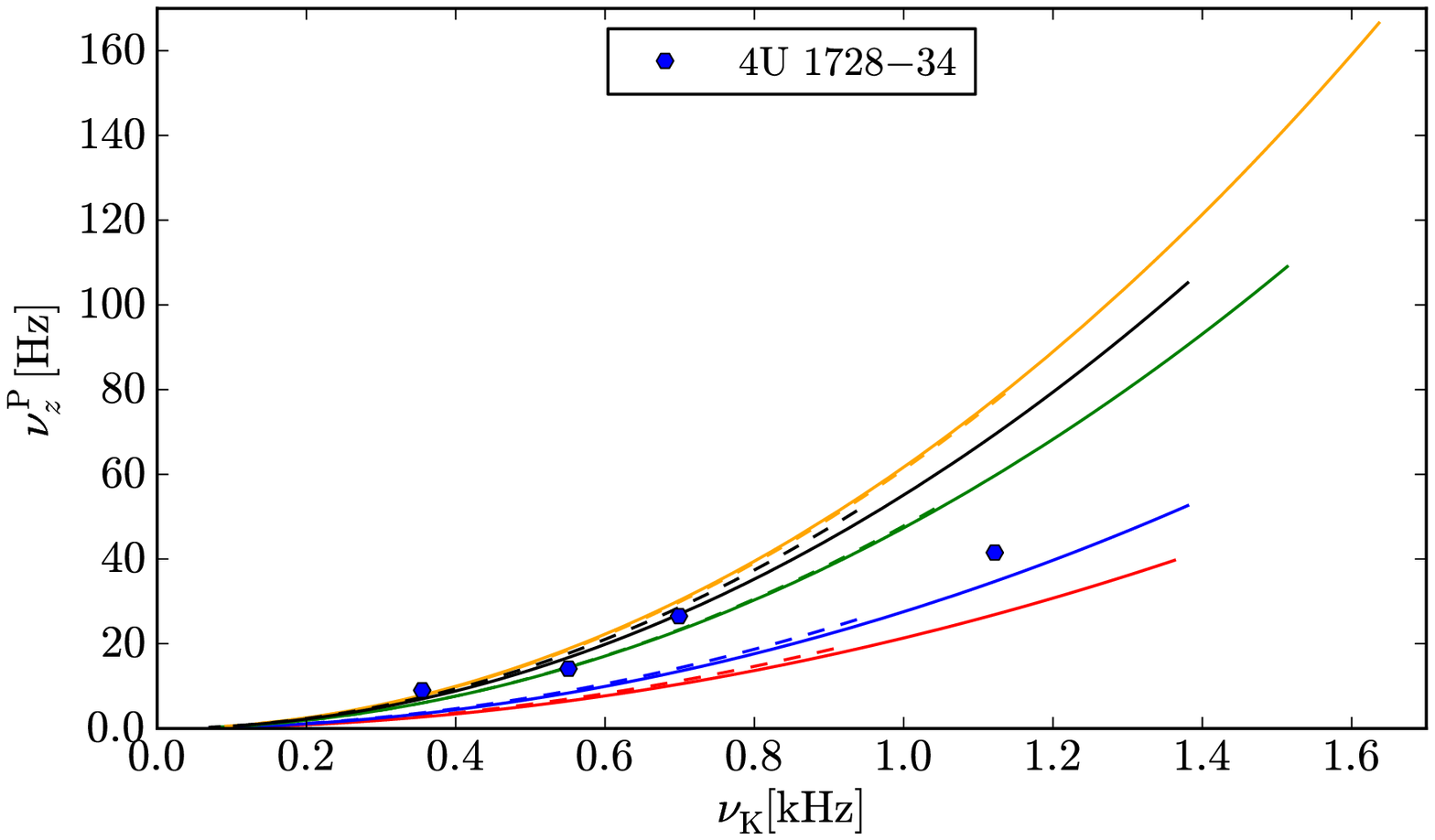}
\caption{Nodal precession frequency, $\nu^{\rm P}_{z}$, as a function of the Keplerian frequency $\nu_{\rm K}$ for the NS realistic models in Table~\ref{tab:FKLTZ}. The convention is as Fig.~\ref{fig:plotomegarhoSMEOSL}. We indicate the QPO frequencies observed in the LMXB 4U 1728--34 \citep[see][]{1998ApJ...506L..39F}. \textcolor{black}{The black curves indicate the $2\nu^{\rm P}_{z}$-$\nu_{\rm K}$ relation for the models M3 and M4 (solid and dashed) following the suggestion of \cite{1999ApJ...524L..63S,1999PhRvL..82...17S}.}}
\label{fig:plotomegarhoSMEOSA}
\end{figure}

In Fig.~\ref{fig:plotomegarhoSMEOSL} we show the relation $\nu^{\rm P}_{z}$ versus $\nu_{\rm K}$ for the models M1--M8 of Table \ref{tab:FKLTZ}. For the sake of comparison we show the low frequency branch observed in the LMXB 4U 1728--34 \citep[see][for details]{1998ApJ...506L..39F}. From the analysis of the pulsating X-ray flux it turns out that very likely the spin frequency of the NS in 4U 1728--34 is $\sim 363$ Hz \citep[see][for details]{1996ApJ...469L...9S}. Thus, the models M3 ($M_0=1.94 M_\odot$, $j=0.24$) and M4 ($M_0=2.71 M_\odot$, $j=0.18$) in Table \ref{tab:FKLTZ} that correspond to a NS of spin frequency $360$ Hz are of particular interest for the analysis of this source. It was suggested by \cite{1999ApJ...524L..63S,1999PhRvL..82...17S} that the low frequency observed in 4U 1728--34 are likely to be due to excitations of the second harmonic of the vertical motion and therefore a better fit of the lower-higher QPO frequencies of 4U 1728--34 (and of similar sources) will be obtained for the relation $2\nu^{\rm P}_{z}$-$\nu_{\rm K}$. The black curves in Fig.~\ref{fig:plotomegarhoSMEOSA} indicate the $2\nu^{\rm P}_{z}$-$\nu_{\rm K}$ relation for the models M3 and M4 (solid and dashed) following the above suggestion. Although the improvement of the fit is evident, we notice that the NS parameters that correctly reproduce the features of 4U 1728--34 are likely in between the models M3 and M4.

\section{Concluding Remarks}\label{sec:6}

We have done an extensive comparison of the orbital motion of neutral test particles in the PRS and Kerr spacetime geometries. In particular we have emphasized on the Keplerian and frame-dragging frequencies, as well as the precession and oscillation frequencies of the radial and vertical motions. 

\textcolor{black}{We have evidentiated the differences in this respect between the Kerr and PRS solution, especially in the rapid $\sim$kHz rotation regime. Such differences are the manifestation of the influence of the high order multipole moments such as the quadrupole and octupole.}

\textcolor{black}{The analysis of the deviations between the Kerr and PRS features for given mass and angular momentum of a source studied in this work are useful to distinguish the signatures between BHs and NSs, which relevant to establish a separatrix for the identification of the compact objects harbored in X-Ray Binaries. In the case of BH candidates, these results might become important for testing the no-hair theorem of BHs \citep[see e.g.][]{2011ApJ...726...11J}. Equally important, the application of the precession and oscillation frequencies to the explanation of QPOs in LMXBs possessing a NS, can unveil information on the NS parameters, leading to a possible identification of the behavior of the nuclear matter EoS at supranuclear densities. In this line, the identification of the rotation frequency of NSs in LMXBs from the pulsating X-ray flux $\nu_{\mathrm{burst}}$, e.g.~the case of 4U 1728--34 \citep{1998ApJ...506L..39F}, 4U 1916--053 \citep{2001ApJ...549L..85G} and more recently the case of IGR J17191--2821 \citep{2010MNRAS.409.1136A}, will certainly help to constrain QPO models as well as the NS parameters.  Additional information coming from the photospheric radius expansion phenomena observed in these systems \citep[see e.g.][for details]{2001ApJ...553L.157M} during their transient activity with Super-Eddington emission can become of paramount importance if combined with the QPO information.}

The generalization of the present work to the electrovacuum case is important to establish the influence of the magnetic dipole and quadrupole moments on the orbital motion of particles around compact objects \citep[see e.g.][]{2010CQGra..27d5001B,2010PhRvD..82l4014S,2012CQGra..29f5012B}.

Interesting effects on the epicyclic frequencies due to the presence of the magnetic dipole have been 
already pointed out recently by \cite{2010CQGra..27d5001B} and \cite{2012CQGra..29f5012B}. 
These effects were predicted after neglecting the contribution of the electromagnetic field to the 
curvature, for $j=0$  see \cite{2010CQGra..27d5001B} and for $j\neq0$ \cite{2012CQGra..29f5012B}. 
\cite{2010CQGra..27d5001B} assumed the model of the star as a dipole magnetic field superposed on a Schwarzschild black hole. In the second work, they studied the case of a magnetized slowly rotating neutron stars; to build the model they  superpose an dipolar magnetic field on the Lense-Thirring geometry. The effects of the magnetic dipole on the location of the ISCO, within the PRS solution, has been investigated by \cite{2010PhRvD..82l4014S}. 

A complete analysis of the effects due to the emergence of electromagnetic structure on the orbital motion of charged particles is therefore of interest and deserve the appropriate attention. Recent observations have shown that for stars with strong magnetic fields the quadrupole and octupole magnetic terms make significant contributions to the magnetic field \citep{2006MNRAS.370..629D}, which indicates that arbitrary higher order multipole components might be required in a realistic model. The presence of a magnetic quadrupole demands the breaking of the reflection symmetry \citep[see] [for details]{2006CQGra..23.3251P}, by means of a slightly change to the Ernst electric potential over the symmetry axis
\begin{align}
\label{PotencialesPRSmod}
\begin{split}
f(z) = \frac{q z^2 + \textrm{i}\mu z + \textrm{i} \zeta}{z^3+z^2(m-\textrm{i}a)-kz+\textrm{i}s}\, ,
\end{split}
\end{align}
a quadrupolar magnetic component $\mathcal{B}_2 = \zeta$ can be introduced to the PRS solution. Such a change generates just a redefinition of the coefficients $f_i$ in Eq.~(\ref{equ:f_i}). In this way the PRS solution can be readily use to explore the effect of strong magnetic fields with non-dipolar structure. 

\acknowledgments
The authors thank Emanuele Berti for directing our attention to the astrophysical relevance of the periastron and nodal precession frequencies. This work was partially supported by Fundaci\'on para la Promoci\'on de la Investigaci\'on y la Tecnolog\'ia del Banco la Rep\'ublica grant number 2879. L.A.P. acknowledges the financial support by the Colombian Institute for the Science and Technology Development (COLCIENCIAS),  and from Comit\'e para el Desarrollo de la Investigaci\'on --CODI-- of Universidad de Antioquia. C.A.V.-T. is supported by Vicerrector\'ia de Investigaciones (UniValle) 
grant number 7859.

\bibliography{rpfApJv2}

\appendix
\section{Metric Functions}\label{app:metricfuncs}

The functions $A$, $B$, $C$, $H$, $G$, $K$, and $I$ used to express the metric functions 
(\ref{eq:metricfuncs}) are given by
\begin{align}
    A &= \displaystyle\sum_{1\leq i < j < k \leq 6} a_{i j\,
    k} r_{i}\,r_{j}\,r_{k}\, ,
\quad
    B =\displaystyle\sum_{
    1\leq i < j \leq 6} b_{i j} r_{i}\,r_{j},
\quad
    C = \displaystyle\sum_{ 1\leq i < j \leq 6} c_{i j}
    r_{i}\,r_{j}\, ,\quad K = \displaystyle\sum_{ 1\leq i < j < k \leq 6} a_{i j\,
    k}\, ,
\\
    H &= z\,A - (\beta_{1} + \beta_{2}+\beta_{3})B+\displaystyle 
    \sum_{ 1\leq i < j < k \leq 6} h_{i j\, k} r_{i}\,r_{j}\,r_{k} +
    \displaystyle\sum_{ 1\leq i < j \leq 6} (\alpha_{i} + \alpha_{j})\,b_{i j}
    \,r_{i}\,r_{j},
\\
    G &= -(\beta_{1} + \beta_{2} + \beta_{3})\,A + z\,B + \displaystyle\sum_{ 1\leq i < j \leq
    6} g_{i j} \,r_{i}\,r_{j} + \displaystyle \sum_{ 1\leq i < j < k \leq 6} (\alpha_{i} +
    \alpha_{j} + \alpha_{k})a_{i j\,k} r_{i}\,r_{j}\,r_{k}, 
\\
    I &= (f_{1} + f_{2} + f_{3})(A - B) + (\beta_{1}+\beta_{2} +
    \beta_{3} - z)\, C + \displaystyle\sum_{ 1\leq i < j < k \leq 6} p_{i j\,k}
    r_{i}\,r_{j}\,r_{k} + \displaystyle\sum_{i=1}^{6} p_{i}\,
    r_{i} + \displaystyle\sum_{ 1\leq i < j \leq 6} [p_{i
    j}-(\alpha_{i} + \alpha_{j})c_{ij}] r_{i}\,r_{j},
\end{align}
with
\begin{align*}
r_i &= \sqrt{\rho^2 + (z-\alpha_i)^2}\, ,\quad 
a_{i j\,k} = (-1)^{i + j + 1}\Lambda_{i j k}\,\Gamma_{l | m n}\, ,\quad 
b_{i j} = (-1)^{i + j}\lambda_{i j}\,H_{l | m n p}\, ,
\\
c_{i j} &= (-1)^{i + j}\lambda_{i j}[f(\alpha_l)\,\Gamma_{m | n p}
- f(\alpha_m)\,\Gamma_{n | p l} + f(\alpha_n)\,\Gamma_{p | l m}
- f(\alpha_p)\,\Gamma_{l | m n}]\, ,
\\
h_{i j\,k} &= (-1)^{i + j + k}\Lambda_{i j k}(e^{*}_1\,\delta_{2 3
| l m n} + e^{*}_2\,\delta_{3 1 | l m n} + e^{*}_3\,\delta_{1 2
| l m n})\, ,\quad
g_{i j} = (-1)^{i + j}\lambda_{i j}(\alpha_{l}\,\Gamma_{m | n p} -
\alpha_{m}\,\Gamma_{n | p l} + \alpha_{n}\,\Gamma_{p | l m} -
\alpha_{p}\,\Gamma_{l | m n})\, ,\\
p_i &= (-1)^i D_{i}[f(\alpha_l)\,H_{m | n p s} - f(\alpha_m)\,H_{n
| p s l} + f(\alpha_n)\,H_{p | s l m} - f(\alpha_p)\,H_{s | l m
n} + f(\alpha_s)\,H_{l | m n p}]\, ,\\
p_{i j} &= (-1)^{i + j} \lambda_{i j}(e^{*}_1\,\Upsilon_{2 3 | l m
n p} + e^{*}_2\,\Upsilon_{3 1 | l m n p} + e^{*}_3\,\Upsilon_{1
2 | l m n p})\, ,\quad
p_{i j\,k} = (-1)^{i + j +1} \Lambda_{i j\,k}(e^{*}_1\,\Psi_{2 3 |
l m n} + e^{*}_2\,\Psi_{3 1 | l m n} + e^{*}_3\,\Psi_{1 2 | l m
n})\, ,\\
\lambda_{i j} &= (\alpha_i - \alpha_j)\,D_i\,D_j\, ,\quad 
\Lambda_{i j\,k} = (\alpha_i - \alpha_j)(\alpha_i -
\alpha_k)(\alpha_j - \alpha_k)\,D_i\,D_j\,D_k\, ,\\
D_i &= \frac{1}{(\alpha_i - \beta_1)(\alpha_i - \beta_2)(\alpha_i -
\beta_3)}\, ,\quad \Gamma_{l | m n} = H_3(\alpha_l)\,\Delta_{1 2 | m n} +
H_3(\alpha_m)\,\Delta_{1 2 | n l} + H_3(\alpha_n)\,\Delta_{1 2
| l m}\, ,
\end{align*}
and
\begin{align*}
\Delta_{l m | n p} &= H_{l}(\alpha_n)\,H_{m}(\alpha_p) -
H_{l}(\alpha_p)\,H_{m}(\alpha_n)\, ,\quad
H_{l}(\alpha_n) = \frac{2 \prod_{p \neq n} (\alpha_p -
\beta^{*}_l)}{\prod_{k \neq l}^{3} (\beta^{*}_l -
\beta^{*}_k)\,\prod_{k = 1}^{3} (\beta^{*}_l - \beta_k)} -
\sum_{k = 1}^{3} \frac{2f^{*}_l\,f_k}{(\beta^{*}_l -
\beta_k)(\alpha_n - \beta_k)}\, ,\\
\delta_{l m | n p s} &= \Delta_{l m | n p} + \Delta_{l m | p s} + \Delta_{l m | s n},\quad
h_{l | m n p} = H_3(\alpha_l)\,\delta_{1 2 | m n p}\, ,\quad
H_{l | m n p} = h_{l | m n p} + h_{m | n p l} + h_{n | p l m} +
h_{p | l m n}\, ,\\
\Psi_{l m | n p s} &= f(\alpha_n)\,\Delta_{l m | p s} +
f(\alpha_p)\,\Delta_{l m | s n} + f(\alpha_s)\,\Delta_{l m | n p}\, ,\\
\Upsilon_{l m | n p r s} &= f(\alpha_n)\,\delta_{l m | p r s} -
f(\alpha_p)\,\delta_{l m | r s n} +f(\alpha_r)\,\delta_{l m | s n
p} - f(\alpha_s)\,\delta_{l m | n p r}\, ,
\end{align*}
being $\alpha$'s the roots of the Sibgatullin equation \citet{1991owsg.book.....S,
1993CQGra..10.1383M}
\begin{equation}\label{eq sibga}
e(z)+\tilde{e}(z)+2{\tilde{f}}(z)f(z)=0.
\end{equation}

\section{Kerr's metric in Weyl-Papapetrou quasi-cylindrical coordinates}
\label{app:Kerrmetricfuncs}
In order to keep comparisons in the save place, we consider useful to display
the Kerr solution in the Weyl-Papapetrou quasi-cylindrical coordinates. For this case,
\begin{align}\label{eq:Kerrmetricfuncs}
f&=\frac{A \bar{A}-B \bar{B}}{( A -
B)(\bar{A}-\bar{B})}\, ,
\quad 
e^{2\gamma}=\frac{A \bar{A} -B
\bar{B} }{\displaystyle{K \bar{K}\prod_{n=1}^{2}r_n}}, \quad
\omega = \frac{{\rm Im}[(A + B)\bar{H}-(\bar{A} + \bar{B})G]}{A \bar{A} - B \bar{B} }\, ,
\end{align}
where for our own convenience we do not present the definition of each term, but present
the final combination of them, i.e.,
\begin{align}
\begin{split}
A \bar{A}-B \bar{B} &=-8 \left(a^2-m^2\right)^3 \left(\rho ^2+z^2\right) \left(m^2 \sqrt{-2 z
   \sqrt{m^2-a^2}-a^2+m^2+\rho ^2+z^2}
   \sqrt{\left(\sqrt{m^2-a^2}+z\right)^2+\rho ^2} \right.
\\   
   &- \left. 2 a^2\sqrt{-2 z
   \sqrt{m^2-a^2}-a^2+m^2+\rho ^2+z^2}
   \sqrt{\left(\sqrt{m^2-a^2}+z\right)^2+\rho ^2}+a^2 m^2-m^4+m^2 \rho ^2+m^2
   z^2\right),
   \end{split}   
\\
\begin{split}
(A -B)(\bar{A}-\bar{B})&=
-8 \left(m^2-a^2\right)^3 
\\
&\times \left(\rho ^2+z^2\right) \left(a^2 \left(2 m \left(\sqrt{-2
   z \sqrt{m^2-a^2}-a^2+m^2+\rho ^2+z^2}+\sqrt{2 z
   \sqrt{m^2-a^2}-a^2+m^2+\rho ^2+z^2}\right)\right.\right.
\\
   &+\left.\left. 2 \sqrt{-2 z
   \sqrt{m^2-a^2}-a^2+m^2+\rho ^2+z^2} \sqrt{2 z \sqrt{m^2-a^2}-a^2+m^2+\rho
   ^2+z^2}+3 m^2\right)\right.
\\   
   &-\left.m^2 \left(2 m \left(\sqrt{-2 z
   \sqrt{m^2-a^2}-a^2+m^2+\rho ^2+z^2}+\sqrt{2 z \sqrt{m^2-a^2}-a^2+m^2+\rho
   ^2+z^2}\right)\right.\right.
\\   
   &+\left.\left.\sqrt{-2 z \sqrt{m^2-a^2}-a^2+m^2+\rho ^2+z^2} \sqrt{2 z
   \sqrt{m^2-a^2}-a^2+m^2+\rho ^2+z^2}+3 m^2+\rho ^2+z^2\right)\right),
\end{split}   
\\
K \bar{K}\displaystyle{\prod_{n=1}^{2}r_n}&= 16 \left(m^2-a^2\right)^4 \left(\rho ^2+z^2\right)
   \sqrt{\left(z-\sqrt{m^2-a^2}\right)^2+\rho ^2}
   \sqrt{\left(\sqrt{m^2-a^2}+z\right)^2+\rho ^2}
\\
\begin{split}
{\rm Im}[(A + B)\bar{H}&-(\bar{A} + \bar{B})G]  = 
16 a m \left(m^2-a^2\right)^3 \left(\rho ^2+z^2\right) \left(-m^2 \sqrt{-2 z
   \sqrt{m^2-a^2}-a^2+m^2+\rho ^2+z^2}
\right. \\&- \left.
m \sqrt{-2 z
   \sqrt{m^2-a^2}-a^2+m^2+\rho ^2+z^2}
   \sqrt{\left(\sqrt{m^2-a^2}+z\right)^2+\rho ^2}
\right. \\&+ \left.   
   a^2 \sqrt{-2 z
   \sqrt{m^2-a^2}-a^2+m^2+\rho ^2+z^2}-z \sqrt{m^2-a^2} \sqrt{-2 z
   \sqrt{m^2-a^2}-a^2+m^2+\rho ^2+z^2}
\right. \\&- \left.
   m^2
   \sqrt{\left(\sqrt{m^2-a^2}+z\right)^2+\rho ^2}+a^2
   \sqrt{\left(\sqrt{m^2-a^2}+z\right)^2+\rho ^2}
   \right. \\&+ \left.
   z \sqrt{m^2-a^2}
   \sqrt{\left(\sqrt{m^2-a^2}+z\right)^2+\rho ^2}+a^2 m-m^3+m \rho ^2+m z^2\right).
\end{split}   
\end{align}
From here, it is clear how changing $a\rightarrow-a$ will cause only a global change in the
sign of the metric function $\omega$ and therefore only a change in the $g_{t\phi}$ metric
component.


\end{document}